\documentclass{llncs}

\usepackage[hyphens]{url}

\usepackage{ifthen}
\newboolean{full}

\usepackage{amsthm}

\usepackage{multicol}
\usepackage{multirow}
\usepackage{amssymb}
\usepackage{amsmath}
\usepackage{geometry}
\usepackage{bbding}
\usepackage{graphicx}
\usepackage{color}
\usepackage{verbatim}
\usepackage{textcomp}

\usepackage[caption=false]{subfig}
\usepackage{float}

\usepackage{listings}

\newcommand{\eg}{e.g.\ }
\newcommand{\ie}{i.e.\ }

\newlength\figwidth
\newcommand{\inserttabd}[4]{%
	\begin{table*}[!t]
		\begin{minipage}{\textwidth}
			\caption{#3}
			\label{#1}
			\centering
			\renewcommand{\arraystretch}{1.15}
			\begin{tabular}{#2}
				\hline
				#4
				\hline
			\end{tabular}
		\end{minipage}
\end{table*}}

\usepackage{hyperref}

\begin{document}
\mainmatter  
%

\title{Anonymous Single-Sign-On for $n$ designated services with traceability}

\titlerunning{}

\author{Jinguang Han, Liqun Chen, Steve Schneider, Helen Treharne and Stephan Wesemeyer}

\authorrunning{Han \em{et. al}}

\institute{Department of
Computer Science, University of Surrey,
Guildford, Surrey, GU2 7XH, United Kingdom\\
}%
%

\toctitle{}
 \tocauthor{} 
 \maketitle
\begin{abstract} Anonymous Single-Sign-On authentication schemes have 
been proposed to allow users to access a service protected by a verifier without
revealing their identity which has become more important due to the introduction
of strong privacy regulations. In this paper we describe a new approach whereby
anonymous authentication to different verifiers is achieved via authorisation
tags and pseudonyms.  %
The particular innovation of our scheme is authentication can only occur between
a user and its designated verifier for a service, and the verification cannot be
performed by any other verifier. %
The benefit of this authentication approach is that it prevents information
leakage of a user's service access information, even if the verifiers for these
services collude which each other. %
Our scheme also supports a trusted third party who is authorised to de-anonymise
the user and reveal her whole services access information if required. %
Furthermore, our scheme is lightweight because it does not rely on attribute or
policy-based signature schemes to enable access to multiple services. The
scheme's security model is given together with a security proof, an
implementation and a performance evaluation. %

\end{abstract}
\medskip

\keywordname{ Anonymous Single-Sign-on, Security, Privacy, Anonymity}

\section{Introduction}
\label{sec:intro}

Single-sign-on (SSO) systems are a user-friendly way of allowing users access to
multiple services without requiring them to have different usernames or
passwords for each service. SSO solutions (\eg OpenID 2.0
\cite{recordon2006openid} by the OpenID foundation or Massachusetts Institute of
Technology (MIT)'s Kerberos \cite{2017MITKerberos}) are designed to make the
users' identities and possibly additional personal identifiable
information~(PII) available to the verifiers of the services which they wish to
access. %
However, for some services, a verifier may not require the user's identity (nor
any associated PII), just that the user is authorised to access the desired
service. %
Moreover, the introduction of more stringent obligations
with regards to the handling of PII in various jurisdictions (\eg GDPR in
Europe\cite{EGDPR2016}), requires service providers to minimise the use of PII. 

Anonymous single-sign-on
schemes\cite{2008ElmuftiAnonAuth,2010hanDynSSO,2013WangAnonSSO,2015LeeAnonSSO}
exist which can protect a user's identity, but may not do so for all entities
within a scheme.   %
Moreover, a user's service request can be verified by all verifiers of a system
rather than the one it is intended for which may pose a potential privacy risk
to both the user and that verifier.


Our proposed scheme addresses these issues and provides the following features:
(1) only one authentication ticket is issued to a user, even if she wants to
access multiple distinct services; (2) a user can obtain a ticket from a ticket
issuer anonymously without releasing anything about her personal identifiable
information --- in particular, the ticket issuer cannot determine whether two
ticket requests are for the same user or two different users; (3) a designated
verifier can determine whether a user is authorised to access its service but
cannot link different service requests made by the same user nor collude with
other verifiers to link a user's service requests; (4) designated verifiers can
detect and prevent a user making multiple authentication requests using the same
authentication tag (``double spend'') but cannot de-anonymise the user as a
result; (5) tickets cannot be forged; and (6) given a user's ticket, a central
verifier is authorised to recover a user's identity as well as the identities of
the verifiers for the requested services in the user's ticket.

Our contributions  are: a novel, anonymous single-sign-on scheme providing the
above features;  %
its associated security model and security definitions; a corresponding formal
proof of its security as well as an empirical performance analysis based on a
Java-based implementation of our scheme.

\subsection{Related Work}

We now look at previous research which is most closely related to our scheme in
the areas of: i) Anonymous Single-Sign-On protocols, ii) anonymous
authentication schemes, iii) multi-coupon schemes and iv) designated verifiers
signature schemes.

\medskip
\noindent{\bf\em Anonymous Single-Sign-On schemes}
\medskip

One of the earliest anonymous single-sign-on system was proposed by Elmufti et
al.\cite{2008ElmuftiAnonAuth} for the Global System for Mobile communication
(GSM). In their system, a user generates a different one-time identity each time
they would like to access a service and, having authenticated the user, a
trusted third party will then authenticate this one-time identity to the
service provider. Consequently, the user is anonymous to the service provider
but, unlike in our scheme, not the trusted third party who authenticated the
one-time identity.

In 2010, Han et al. \cite{2010hanDynSSO} proposed a novel dynamic SSO 
system which uses a digital signature to guarantee both the 
unforgeability and the public verification of a user's credential. In 
order to protect the user's privacy, their scheme uses broadcast 
encryption which means that only the designated service providers can 
check the validity of the user's credential. Moreover, zero-knowledge 
proofs are used to show that the user is the owner of those valid 
credentials to prevent impersonation attacks. However, again unlike our 
scheme, the user is still known to the trusted third party which issued 
the credentials.

Wang et al. \cite{2013WangAnonSSO}, on the other hand, propose an anonymous SSO
based on group signatures \cite{bmw:gs2003}. In order to access a service, the
user generates a different signature-based  pseudonyms from her credentials and
sends the signature to the service provider. If the signature is valid, the
service provider grants the user access to the service to the user; otherwise,
the service request is denied. The real identities of users can be identified by
using the opening technique in \cite{bmw:gs2003}. While the user remains
anonymous, their scheme (unlike ours) does not, however, provide designated
verifiers, \ie all verifiers can validate a user's request.

Lastly, Lee\cite{2015LeeAnonSSO} proposed an efficient anonymous SSO based on
Chebyshev Chaotic Maps. In this scheme, an issuer, the ``smart card processing
center'', issues secret keys to users and service providers when they join in
the system and to access a service, a user and service provider establish a
session key with their respective secret keys. If the session key is generated
correctly, the service request is granted; otherwise, it is denied. However, 
unlike our scheme, each service provider knows the identity of
the user accessing their service.

While in \cite{2010hanDynSSO}, \cite{2013WangAnonSSO} and \cite{2015LeeAnonSSO},
a user can access any service in the system by using her credentials, in our
scheme, a user can only access the services which she selects when obtaining a
ticket but can do so while remaining completely anonymous to both issuer and 
service provider.

\medskip
\noindent{\bf\em Anonymous authentication schemes}
\medskip

With respect to anonymous authentication solutions, we consider schemes whose
primary feature is to support multiple anonymous authentication. As in our
scheme, anonymous authentication enables users to convince verifiers that they
are authorised users without releasing their exact identities. %

Teranishi {\em et al.} \cite{tfs:k-taa04} proposed a $k$-times anonymous
authentication ($k$-TAA) scheme where the verifiers determine the number of
anonymous authentication that can be performed.  The $k$-TAA scheme provides the
following two features: (1) no party can identify users who have been
authenticated within $k$ times; (2) any party can trace  users who have been
authenticated more than $k$ times. The verifier generates $k$ tags and for each
authentication, a user selects a fresh tag. Nguyen {\em et al.}
\cite{nr:k-taa05} proposed a similar dynamic $k$-TAA scheme to restrict access
to services not only the number of times but also other factors such as expiry
date.

Camenisch {\em et al.} \cite{chklm:k-taa06} proposed a periodic $k$-TAA 
scheme which enables users to authenticate themselves to the verifiers 
no more than $k$ times in a given time period but supports reuse of the 
$k$ times authentication once the period is up. In this scheme, the 
issuer decides the number of anonymous authentication request a user 
can make in a given time period. When a user makes an anonymous 
authentication request, he proves to a verifier that he has obtained a 
valid CL signature from the issuer. 

Note, however, that our scheme also prevents a verifier from establishing
whether a user has used any of the other services thereby also guaranteeing
verifier anonymity.

Furthermore, in all of these $k$-TAA schemes
\cite{tfs:k-taa04,nr:k-taa05,chklm:k-taa06}, authentication is not bound to a
particular verifier, whereas in our scheme authentication tags are bound to
specific verifiers. Moreover, $k$-TAA schemes allow verifiers to determine a
user's identity who has authenticated more than $k$ times while in our scheme
multiple authentications to a single verifier is considered ''double spending''
which a verifier can detect but which does not lead to the de-anonymisation of a
user.

However, to prevent users from potentially abusing the system,  our scheme
allows for a central verifier who, given a user's ticket, can extract from it
both the user's and verifiers' public keys using the authentication tags
contained within it and thus establish the identities of both the user and her
associated verifiers.

Lastly, Camenisch {\em et al.} in \cite{camenisch2010} and the IBM identity
mixer description of its features in \cite{ibmmixer} define a scheme that has
similar properties to ours including that of a central verifier (called
``inspector'') trusted to reveal a user's identity. The scheme is based on users
obtaining a list of certified attributes from an issuer and the users using a
subset of their attributes to authenticate to verifiers. The distinguishing
difference between their scheme and ours is that their verification of anonymous
credentials is not bound to a designated verifier whereas our is.



\medskip
\noindent{\bf\em Multi-coupon schemes}
\medskip

There is some degree of similarity between our scheme and a number of multi-coupon
schemes.  Armknecht {\em et al.} \cite{aelms:mc2008} proposed a multi-coupon
scheme for federated environments where multiple vendors exist. In
\cite{aelms:mc2008},  a user can redeem  multiple coupons anonymously with
different vendors in an arbitrary order. To prevent double-spending of a coupon,
a central database is required to record the transaction of each multi-coupon.
The main difference to our scheme is that each coupon can be redeemed against
any service provider while our authentication tags can only be validated by its
designated verifier. Moreover, our ``double-spend'' detection is done by the
verifier and does not require a central database.

Similarly, the schemes propose by Liu {\em et al.} \cite{lmyy:2017} which
provides strong user privacy and where a user can use an e-coupon anonymously no
more than $k$ times before his identity can be recovered. However, the user's 
coupons can be redeemed against any service rather than a designated verifier 
as our scheme provides.

\medskip
\noindent{\bf\em Designated Verifiers}
\medskip

Jakobsson in \cite{jakobsson_designated09} introduced the concept of a
designated verifier which means that in a proof we ascertain that nobody but
this verifier can be convinced by that proof while the authors in
\cite{FAN2012944} present an anonymous attribute based scheme using
designated-verifiers. In their work they focus on identifying multiple
designated verifiers This is achieved through using the verifier's private key
in the verification so that no other third party can validate the designated
verifier signature. We adopt the high level concept of a designated verifier in
our approach, \ie given a valid authentication tag for service $A$, only service
$A$'s verifier can establish its validity. As this property is conceptually
similar to the designated signatures described in
\cite{jakobsson_designated09,FAN2012944}, our verifiers are called designated
verifiers. However, this is where the similarity ends with Jakobsson's
designated verifiers. Notably, in \cite{jsi:1996}, a verifier cannot convince
others that the signature is from the signer because the verifier can generate
the signature by himself. In our scheme, everyone can check that the
authentication tags are signatures generated by the ticket issuer.

In summary, while a number of previous authentication schemes address the
anonymity of the user and multiple authentications, the novelty of our work is
that we ensure no information leakage across verifiers, since authentication can
only occur between a user and its designated verifier while also providing a
central verifier who can de-anonymise the user and reveal the identity of the
verifiers in case of a misbehaving user. %
To the best of our knowledge, our anonymous single-sign-on scheme using
designated verifiers is the first which has been formally presented in term of
definitions, security models and proven to be secure under various cryptographic
complexity assumptions together with an empirical performance evaluation.

\subsection{Paper Organisation}%

This paper is organised as follows: Section~\ref{sec:contributions} provides a
high-level overview of the scheme and its claimed security properties;
Section~\ref{sec:secmodelsum} outlines the applicable security model; Section
\ref{sec:preliminaries} introduces the mathematical concepts and notation used
throughout this paper; Section~\ref{sec:scheme} describes the mathematical
construction of our while Section~\ref{sec:security} presents the theorems for
its security proof; Section~\ref{sec:benchmarks} provides a performance
evaluation of our scheme; and Section~\ref{sec:conclusion} concludes the paper
with directions for future work. 

\section{Scheme overview and security properties}
\medskip%
\noindent{\sf\bf Entities in our proposed scheme}%
\medskip%

\begin{figure*}[t]
\centering
\includegraphics[width=10cm,height=6cm]{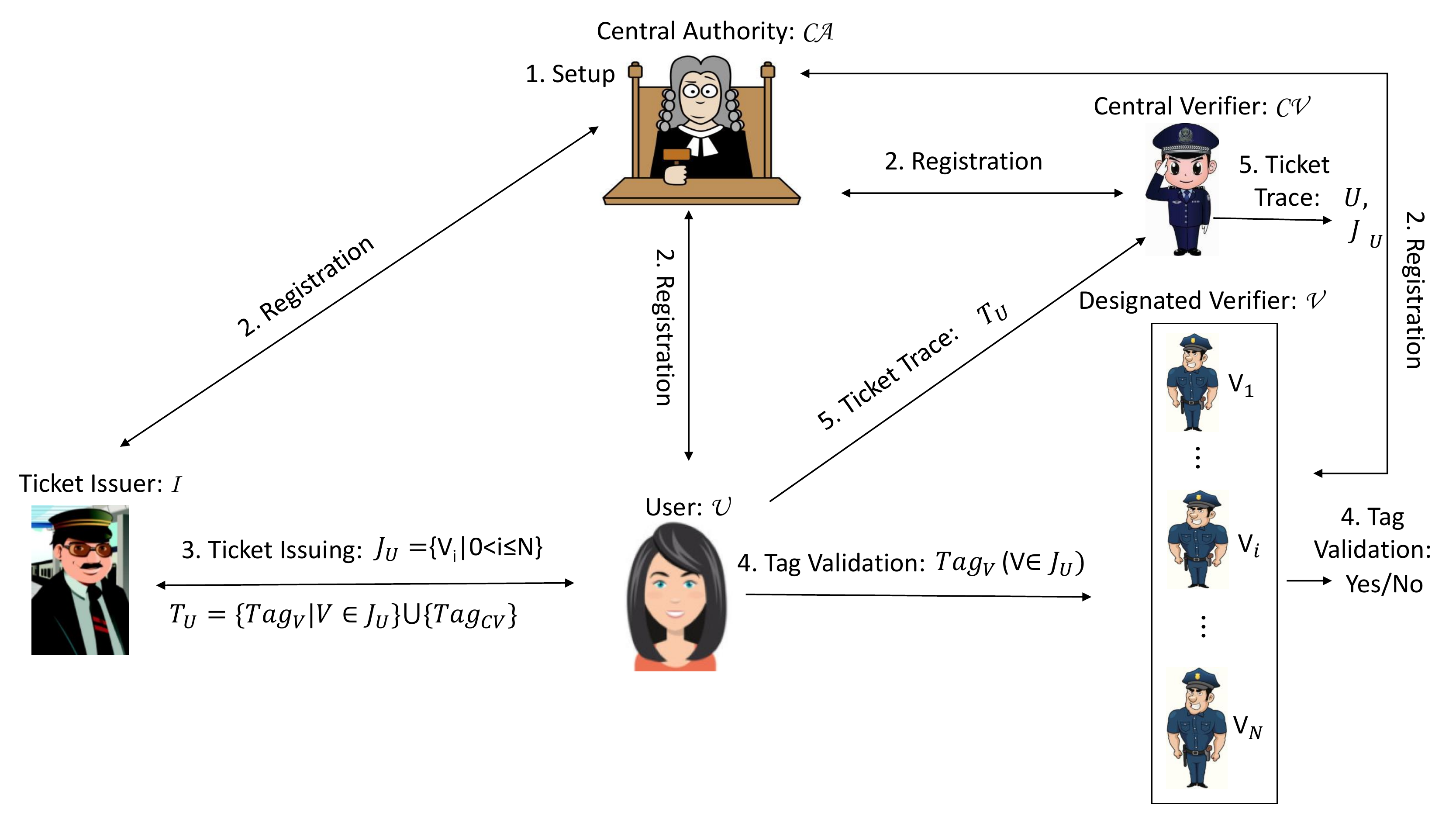}
\caption{Interaction of the various entities in our scheme}\label{fig:HLInteraction}
\end{figure*}

Before providing a high-level overview of our anonymous single-sign-on scheme,
we first introduce the various entities in the scheme as
shown in Figure \ref{fig:HLInteraction}, and define their purpose and roles:

\begin{itemize}

\item {\bf Central Authority ($\mathcal{CA}$):} The $\mathcal{CA}$ is a trusted
third party responsible for establishing the cryptographic keys and parameters
used in the scheme and signing the public keys of the other entities in the
scheme;

\item {\bf User ($\mathcal{U}$):} Someone who wishes to access some distinct
services anonymously;

\item {\bf Ticket Issuer ($\mathcal{I}$):} This entity issues tickets to
registered, yet anonymous users for the requested services;

\item {\bf Designated Verifier ($\mathcal{V}$):} The $\mathcal{V}$ is a verifier
for a specific service that a user might want to access;

\item {\bf Central Verifier ($\mathcal{CV}$):} $\mathcal{CV}$  is another
trusted third party which is allowed to retrieve the identities of the user,
$\mathcal{U}$, and the verifiers, $\mathcal{V}$s, from the authentication tags
present in a user's ticket, $\mathcal{T_U}$.

\item {\bf Authentication Tag ($Tag_V$):} This tag is both tied to a user,
$\mathcal{U}$, and a designated verifier, $\mathcal{V}$ and is used to prove 
to the designated verifier that the user is a valid user and allowed to access 
the associated service;

\item {\bf Ticket ($\mathcal{T_U}$):} A ticket which contains the
authentication tags for the services a user, $\mathcal{U}$, has requested;

\end{itemize}

\medskip%
\noindent{\sf\bf Overview of proposed scheme}%
\medskip%

Figure~\ref{fig:HLInteraction} illustrates at a high-level how our scheme works.
For the detailed mathematical construction of our scheme,  please refer to
Section~\ref{sec:scheme}. Conceptually, our scheme operates as follows:
\begin{itemize}%
	
\item{\sf\bf Registration:} The issuer, verifiers, central verifier and
users all register with the CA.
	
\item{\sf\bf Ticket Issuing:} A user decides which services(and thus which
verifiers) she wants to access and requests an appropriate ticket from the
issuer.
	
\item{\sf\bf Tag Validation:} To access a service, the user presents the
appropriate authentication tag to the service. The validity period and any other
restrictions of the tag can be captured in the free text part of the tag or be a
default set by the verifier. If a user's tag is valid then the user is logged in
to the service. Note that, unlike some other Single-Sign-On systems, the issuer
does not need to be on-line for the tag validation to succeed.
	
\item{\sf\bf ``Double-Spend'' detection:} Should the user present the same tag
twice then the verifier can warn the user that she is already logged in and
should resume the already existing session or offer to terminate the previous
session and continue with a fresh one.
	
\item{\sf\bf Ticket trace:} If a user is seen to abuse the service (\eg violate
the terms and conditions), the central verifier might be called upon to
de-anonymise the user and determine any other services she has used.
\end{itemize}

\medskip%
\noindent{\sf\bf Security properties in our proposed scheme}%
\medskip%

Having defined the different entities and described how they interact, we now 
list the security properties of our scheme:

\begin{itemize}

\item {\sf\bf User Anonymity}: In our scheme, users use pseudonyms whenever
they interact with the issuer or a verifier. As such, the issuer cannot link a
user across different ticket requests. Similarly, a user's identity is also 
hidden from a designated verifier.

\item {\sf\bf Authentication Tag Unlinkability:} Apart from the central verifier
and the issuer, no set of colluding verifiers can establish whether two or more
different authentication tags came from the same anonymous user.

\item {\sf\bf Verifier Anonymity}: The verifier's identify is protected from 
other
users and verifiers, \ie given an authentication tag, only the designated
verifier can validate it and no other verifier (apart from the central verifier
and the issuer) can determine for whom it is.

\item {\sf\bf Designated Verifiability}: Given an authentication tag, $Tag_V$ 
for
verifier, $\mathcal{V}$, only $\mathcal{V}$ can validate it. 

\item {\sf\bf ``Double-spend'' detection}: Any verifier, $\mathcal{V}$, can 
detect
when a user attempts to re-use an authentication tag but cannot de-anonymise the
user.

\item {\sf\bf Unforgeability}: Neither tickets nor individual authentication 
tags
can be forged by any colluding users or verifiers.

\item{\sf\bf Traceability: }There exists a trusted third party, a central 
verifier, who can, given a user's ticket, $T_U$, retrieve the user's and the 
verifiers' public keys (and hence their respective identities) from the 
authentication tags contained within $T_U$.
\end{itemize}

In the next section, we provide the security models in which these properties
hold while Section~\ref{sec:security} contains the associated theorems which are
used to prove those models.

\section{Security Model Overview}
\label{sec:secmodelsum}
We now present a high-level overview of the security models which are used to
prove the security of our scheme. The models are defined by the following games
executed between a challenger and an adversary.
\medskip 

\noindent{\sf\bf Unlinkability Game:} 
\medskip

\noindent This game covers the security properties of user anonymity,
authentication tag unlinkability, verifier anonymity, designated verifiability
and ``double spend'' detection. In this game verifiers and other users can
collude but cannot profile a user's whole service information. In other words,
no party can  link different tags to the same user and determine a verifier's
identity included in an authentication tag (thus proving verifier anonymity)
except for the designated verifier, the ticket issuer or the central verifier.
Moreover, for each authentication tag, the adversary can query its validity once,
which in the context of this game addresses the properties of designated
verifiability and ``double spending''.

This game is formally defined in Appendix \ref{unlink-game}.
\medskip

\noindent{\sf\bf Unforgeability Game:}
\medskip

\noindent This game focuses on proving the unforgeability property of our
scheme. Users, verifiers and the central verifier can collude but cannot forge a
ticket on behalf of the ticket issuer.

This game is formally defined in Appendix 
\ref{unforge-game}.

\medskip
\noindent{\sf\bf Traceability Game:} 
\medskip

\noindent This game focuses on the traceability property of our scheme. It shows
that even if users, verifiers and the central verifier collude, they cannot
generate a ticket which is linked to a user who has never obtained a ticket or a
user who is not the real owner of the ticket.

This game is formally defined in Appendix 
\ref{trace-game}.
\medskip

\section{Preliminaries}\label{sec:preliminaries}

In this section, we introduce the mathematical concepts used by our scheme
including bilinear groups, the BBS+ signature scheme, zero knowledge proofs and
various complexity assumptions needed to ensure its security. The mathematical
notation and symbols used throughout this paper are summarised in Table
\ref{syntax}.

\begin{table*}[t]\caption{Syntax Summary}\label{syntax}
\centering
\begin{tabular}{|l|l|l|l|}
\hline
Syntax & ~~~~~~~~~Explanations & Syntax &~~~~~~~~~~~~~~~~~ Explanations \\
\hline
$1^{\ell}$   ~             &A security parameter  &$V_{i}$ & The $i$-th ticket 
verifier\\
$\mathcal{CA}$ ~    &Central authority &$J_{U}$             ~    &The service 
set of $\mathcal{U}$ consisting of the identities \\
$\mathcal{I}$ ~       &Ticket issuer  &  &of ticket verifiers \& $ID_{CV}$ \\
$\mathcal{V}$   ~     &Ticket verifier &$PP$ & Public parameters\\
$\mathcal{U}$     ~   &User   & $Ps_{U}$               &A set of pseudonyms of  
$\mathcal{U}$ \\
$\mathcal{CV}$~        &Central verifier  &$Ps_{V}$     ~     & The  pseudonym 
generated for the verifier $\mathcal{V}$ \\
$ID_{I}$          ~      &The identity of $\mathcal{I}$ &$Tag_{V}$            
~   &  An  authentication tag for the verifier $\mathcal{V}$\\
$ID_{V}$              ~  &The identity of $\mathcal{V}$ ~  
&$Tag_{CV}$             &   An authentication tag for $\mathcal{CV}$\\
$ID_{U}$                ~&The identity of $\mathcal{U}$  ~&$T_{U}$             
~  & A ticket issued to $\mathcal{U}$\\
$ID_{CV}$                ~&The identity of $\mathcal{CV}$  ~& 
$|X|$                    ~& The cardinality of the set $X$  \\
$\epsilon(\ell)$       ~&A negligible function in $\ell$  ~& 
$x\stackrel{R}{\leftarrow} X$ ~&$x$ is randomly selected from the set $X$\\
$\sigma_{I}$         ~&The credential of $\mathcal{I}$   ~& $A(x)\rightarrow y$ 
~&$y$ is computed by running the algorithm $A(\cdot)$ \\
$\sigma_{V}$         ~&The credential of $\mathcal{V}$   ~& & with input $x$\\
$\sigma_{U}$        ~&The credential of $\mathcal{U}$    ~& 
$\mathcal{KG}(1^{\ell})$ ~&A secret-public key pair generation algorithm\\
$\sigma_{CV}$     ~    &The credential of $\mathcal{CV}$   ~& 
$\mathcal{BG}(1^{\ell})$ ~&A bilinear group generation algorithm\\
$MSK$ &Master Secret Key &$p,q$ & prime numbers\\
$H_1, H_2$ &cryptographic hash functions &&\\
\hline
\end{tabular}
\end{table*}

\subsection{Bilinear Groups and pairings}
In our scheme, bilinear groups are used to support the BBS+ signature scheme 
(defined in Section~\ref{sec:bbs} below).

Let $\mathbb{G}_{1}$, $\mathbb{G}_{2}$ and $\mathbb{G}_{\tau}$ be three cyclic
groups with prime order $p$. A  pairing is defined to be a bilinear,
non-degenerative and computable map $e:\mathbb{G}_{1}\times
\mathbb{G}_{2}\rightarrow \mathbb{G}_{\tau}$\cite{bf:2001}. Given a security
parameter, $1^{\ell}$, we define
$\mathcal{BG}(1^{\ell})\rightarrow(e,p,\mathbb{G}_{1},\mathbb{G}_{2},
\mathbb{G}_{\tau})$ to be a bilinear group generation algorithm. %
Note that Galbraith, Paterson and Smart \cite{gps:2008} classified parings into
three basic types and our scheme is based on the Type-III pairing where the
elements on $\mathbb{G}_{1}$ are short ($\approx 160$ bits). This was chosen
because for all $g\in\mathbb{G}_{1}$ and $\mathfrak{g}\in\mathbb{G}_{2}$, there
exists an polynomial-time efficient algorithm to compute
$e(g,\mathfrak{g})\in\mathbb{G}_{\tau}$ resulting in an more efficient 
algorithm.

S


\subsection{BBS+ Signature}%
\label{sec:bbs}%
Based on the group signature scheme \cite{bbs:2004}, Au, Susilo and Mu
\cite{asm:2006} proposed the  BBS+ signature. This signature scheme works as
follows:

\begin{itemize}%
\item{\sf Setup:} Let
$\mathcal{BG}(1^{\ell})\rightarrow(e,p,\mathbb{G}_{1},\mathbb{G}_{2},
\mathbb{G}_{\tau})$,  $h$ be a generator of $\mathbb{G}_{1}$ and
$g,g_{0},g_{1},\cdots,g_{n}$ be generators of $\mathbb{G}_{2}$. %
\item{\sf KeyGen:} The signer selects $x\stackrel{R}{\leftarrow}\mathbb{Z}_{p}$
and computes $Y=h^{x}$. The secret-public key pair is $(x,Y)$. %
\item{\sf Signing:} To sign a block message $(m_{1},m_{2},\cdots,m_{n})\in
\mathbb{Z}_{p}^{n}$, the signer selects
$w,e\stackrel{R}{\leftarrow}\mathbb{Z}_{p}$, and computes
$\sigma=(g_{0}g^{w}\prod_{i=1}^{n}g_{i}^{m_{i}})^{\frac{1}{x+e}}$. This
signature on $(m_{1},m_{2},\cdots,m_{n})$ is $(w,e,\sigma)$. %
\item{\sf Verification:}  Given a signature $(w,e,\sigma)$ and
$(m_{1},m_{2},\cdots,m_{n})$, the verifier checks
$e(Yh^{e},\sigma)\stackrel{?}{=}e(h,g_{0}g^{w}\prod_{i=1}^{n}g_{i}^{m_{i}})$.
If so, the signature is valid; otherwise, it is invalid. %
\end{itemize}

Au, Susilo and Mu \cite{asm:2006}  reduced the security of the above signature
to the $q$-SDH assumption (see Definition~\ref{def:qSDH} below) in Type-II
paring. Recently, Camenisch,  Drijvers and Lehmann \cite{cdl:daa}  reduced its
security to the JOC-version-$q$-SDH assumption (see Definition~\ref{def:jocqSDH}
below) for Type-III pairing.

\subsection{Zero-Knowledge Proof}

In our scheme, zero-knowledge proof of knowledge protocols are used to 
prove knowledge and statements about various discrete logarithms 
including: (1) proof of knowledge of a discrete logarithm modulo a 
prime number \cite{sch:1991}; (2) proof of knowledge of equality of 
representation \cite{cp:1992}; (3) proof of knowledge of a commitment 
related to the product of two other commitments \cite{cm:1999}. We 
follow the definition  introduced by Camenish and Stadler in 
\cite{cs:1997} which was formalised by Camenish, Kiayias and Yung in 
\cite{cky:2009}. By 
$\mbox{PoK:}\{(\alpha,\beta,\gamma): 
\Upsilon=g^{\alpha}h^{\beta}~\wedge~\tilde{\Upsilon}=\tilde{g}^{\alpha} 
\tilde{h}^{\gamma}\},$ 
proof on knowledge of integers $\alpha$ $\beta$ and $\gamma$ such that 
$\Upsilon=g^{\alpha}h^{\beta}$ and 
$\tilde{\Upsilon}=\tilde{g}^{\alpha}\tilde{h}^{\beta}$ hold on the 
groups $\mathbb{G}=\langle g \rangle=\langle h \rangle$ and 
$\tilde{\mathbb{G}}=\langle \tilde{g} \rangle=\langle \tilde{h} 
\rangle$, respectively.  The convention is that the letters in the 
parenthesis $(\alpha,\beta,\gamma)$ represent the knowledge which is 
being proven by using the other values which the verifier can have 
access to.

\subsection{Complexity Assumptions}
The security of our scheme relies on a number of complexity assumptions defined 
in this subsection.

\begin{definition}{\sf (Discrete Logarithm (DL) Assumption \cite{g:dl1993})}
Let $\mathbb{G}$ be a cyclic group with prime order $p$ and $g$ be a generator of $\mathbb{G}$. Given $Y\in \mathbb{G}$, we say that the discrete logarithm (DL) assumption holds on $\mathbb{G}$ if for all adversary can output a number $x\in\mathbb{Z}_{p}$ such that $Y=g^{x}$ with a negligible advantage, namely 
\begin{equation*}
Adv_{\mathcal{A}}^{DL}=\Pr\left[Y=g^{x}|\mathcal{A}(p,g,\mathbb{G},Y)\rightarrow
 x\right]\leq \epsilon(\ell).
\end{equation*}
\end{definition}
The $DL$ assumption is used in the proof of the traceability property of our
scheme.

\begin{definition}%
\label{def:qSDH}%
{(\sf $q$-Strong Diffie-Hellman ($q$-SDH ) Assumption \cite{bb:2004})} Let
$\mathcal{BG}(1^{\ell})\rightarrow(e,p,\mathbb{G}_{1},\mathbb{G}_{2},
\mathbb{G}_{\tau})$. Suppose that  $g$ and $\mathfrak{g}$ are generators of
$\mathbb{G}_{1}$ and $\mathbb{G}_{2}$, respectively. Given a $(q+2)$-tuple
$(g,g^{x},g^{x^{2}},$ $\cdots,g^{x^{q}},\mathfrak{g})\in
\mathbb{G}_{1}^{q+1}\times \mathbb{G}_{2}$, we say that $q$-strong
Diffie-Hellman assumption holds on
$(e,p,\mathbb{G}_{1},\mathbb{G}_{2},\mathbb{G}_{\tau})$ if for all probabilistic
polynomial-time (PPT) adversary $\mathcal{A}$ can output
$(c,g^{\frac{1}{x+c}})\in\mathbb{Z}_{p}\times\mathbb{G}_{1}$ with a
negligible advantage, namely
\begin{equation*}
Adv_{\mathcal{A}}^{q-SDH}=\Pr\left[\mathcal{A}(\mathfrak{g},g,g^{x},g^{x^{2}},
\cdots,g^{x^{q}})\rightarrow(c,g^{\frac{1}{x+c}})\right]\leq \epsilon(\ell),
\end{equation*} where $c\in\mathbb{Z}_{p}-\{-x\}$. 
\end{definition}

\begin{definition}%
\label{def:jocqSDH}
{\sf ((JOC Version) $q$-Strong Diffie-Hellman (JOC-$q$-SDH) Assumption
\cite{bb:2008})} Let
$\mathcal{BG}(1^{\ell})\rightarrow(e,p,\mathbb{G}_{1},\mathbb{G}_{2},$
$\mathbb{G}_{\tau})$. Given a $(q+3)$-tuple
$(g,g^{x},\cdots,g^{x^{q}},\mathfrak{g},
\mathfrak{g}^{x})\in\mathbb{G}_{1}^{q+1}\times\mathbb{G}_{2}^{2}$, we say that
the JOC- $q$-strong Diffie-Hellman  assumption holds on the bilinear group
$(e,p,\mathbb{G}_{1},\mathbb{G}_{2},\mathbb{G}_{\tau})$ if for all probabilistic
polynomial-time (PPT) adversaries $\mathcal{A}$ can outputs
$(c,g^{\frac{1}{x+c}})\in\mathbb{Z}_{p}\times\mathbb{G}_{1}$ with a negligible
advantage, namely 
\begin{equation*}Adv_{\mathcal{A}}^{JOC-q-SDH}=
\Pr\left[(c,g^{\frac{1}{x+c}})\leftarrow\mathcal{A}(g,g^{x},\cdots,g^{x^{q}},
\mathfrak{g},\mathfrak{g}^{x})\right]<\epsilon(\ell),
\end{equation*} 
where $c\in\mathbb{Z}_{p}-\{-x\}$.
\end{definition}

The security of the BBS+ signature used in our scheme relies on both the
($q$-SDH ) and JOC-$q$-SDH) assumptions.

\begin{definition}%
{\sf (Decisional  Diffie-Hellman (DDH) Assumption \cite{dh:1976})} Let
$\mathcal{BG}(1^{\ell})\rightarrow(e,p,\mathbb{G}_{1},\mathbb{G}_{2},$
$\mathbb{G}_{\tau})$. Give a 3-tuple
$(\xi,\xi^{\alpha},\xi^{\beta},T)\in\mathbb{G}_{1}^{3}$, we say that the
decisional Deffie-Hellman assumption holds on
$(e,p,\mathbb{G}_{1},\mathbb{G}_{2},\mathbb{G}_{\tau})$ if for all probabilistic
polynomial-time (PPT) adversaries $\mathcal{A}$ can distinguish
$T=\xi^{\alpha\beta}$  or $T=M$ with negligible advantage,
namely %
\begin{equation*} 
Adv_{\mathcal{A}}^{DDH}=|\Pr[\mathcal{A}(\xi,\xi^{\alpha},\xi^{\beta},
T=\xi^{\alpha\beta})=1]
-\Pr[\mathcal{A}(\xi,\xi^{\alpha},\xi^{\beta},T=M)=1]|<\epsilon(\ell)
\end{equation*} where $M\stackrel{R}{\leftarrow}\mathbb{G}_{1}$. 
\end{definition}

Note that the DDH assumption is believed to be hard in both $\mathbb{G}_{1}$ and
$\mathbb{G}_{2}$ for the Type-III pairing \cite{gsw:2010} used in our scheme
which means that we actually makes use of the following stronger complexity
assumption.

\begin{definition}%
{\sf (Symmetric External  Diffie-Hellman (SXDH) Assumption \cite{gsw:2010})} Let
$\mathcal{BG}(1^{\ell})\rightarrow(e,p,\mathbb{G}_{1},$
$\mathbb{G}_{2},\mathbb{G}_{\tau})$. We say that the symmetric external
Diffie-Hellman assumption holds on $(e,p,\mathbb{G}_{1},\mathbb{G}_{2},$
$\mathbb{G}_{\tau})$ if the decisional Diffie-Hellman (DDH) assumption holds on
both $\mathbb{G}_{1}$ and $\mathbb{G}_{2}$. %
\end{definition}

\section{Scheme construction}%
\label{sec:scheme}

In this section, we present a more detailed description of the interactions (cf.
Fig. \ref{fig:HLInteraction}) between the entities of our scheme. These
interactions are: (i) System Initialisation, (ii) Registration, (iii) Ticket
Issuing, (iv) Tag Verification and (v) Ticket Tracing. Moreover, we provide
details of the mathematical constructs used in these interactions.
\ifthenelse{\boolean{full}}{%
The formal definitions of these algorithms presented in this section can be
found in Appendix~\ref{app:formal_def_alg}.}%
{Formal definitions of the algorithms presented in this section can be found in
Appendix~\ref{app:formal_def_alg}.}

 

\subsection{System Initialisation}

\begin{figure}[!h]
\centering
\fbox{
\begin{minipage}{14cm}

\noindent{\sf System Set-up:}%
$\mathcal{CA}$ runs $\mathcal{BG}(1^{\ell})\rightarrow(e,p,\mathbb{G}_{1},
\mathbb{G}_{2},\mathbb{G}_{\tau})$ with
$e:\mathbb{G}_{1}\times\mathbb{G}_{2}\rightarrow\mathbb{G}_{\tau}$. Let
$g,h,\xi,\tilde{h}$ be  generators of the group $\mathbb{G}_{1}$ and
$\mathfrak{g}$ be generators of $\mathbb{G}_{2}$. Suppose that
$H_{1}:\{0,1\}^{*}\rightarrow\mathbb{Z}_{p}$ and
$H_{2}:\left\{0,1\right\}^{*}\rightarrow \mathbb{Z}_{p}$  are two cryptographic
hash functions. $\mathcal{CA}$ selects
$x_{a}\stackrel{R}{\leftarrow}\mathbb{Z}_{p}$ and computes
${Y}_{A}=\mathfrak{g}^{x_{a}}$. The master secret key is $MSK=x_{a}$ and the
public parameters are $PP=(e,p,\mathbb{G}_{1},\mathbb{G}_{2},\mathbb{G}_{\tau},
g,h,\xi,\tilde{h},\mathfrak{g},Y_{A},H_{1},H_{2})$. %
\end{minipage}
}\caption{System Set-up Algorithm}\label{fig.setup}
\end{figure}

Fig. \ref{fig.setup} shows the details of the system initialisation in which the
central authority $\mathcal{CA}$ generates a master secret key, $MSK$, and the
required public parameters, $PP$. 

{\bf Note:} Once the system has been set up, all communication between the
different entities in our scheme is assumed to be over secure, encrypted
channels which can be established by the various entities using standard Public
Key Infrastructure. This ensures that our scheme is not susceptible to simple 
Man-In-The-Middle attacks. 

\subsection{Registration}

\begin{figure}[!h]
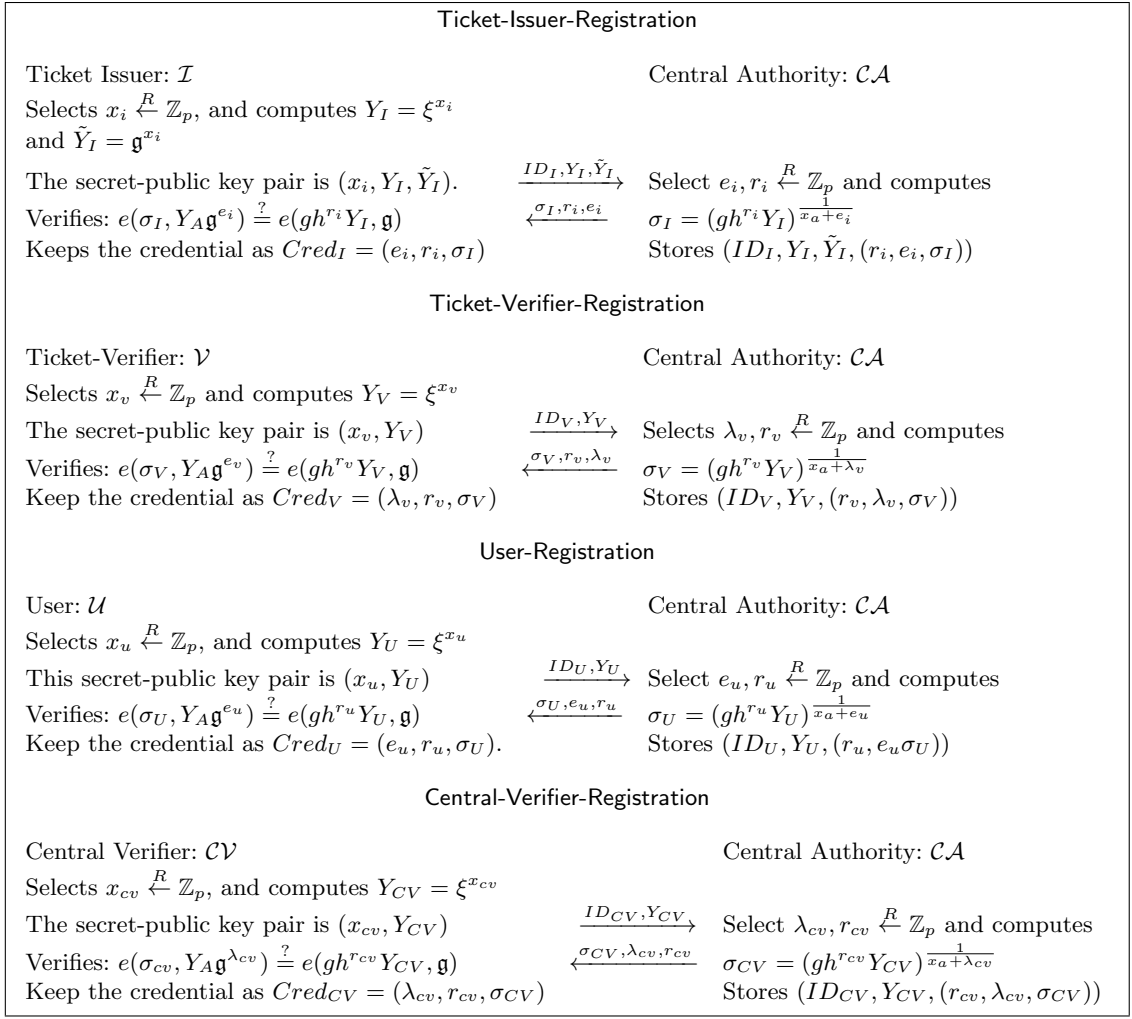

\centering
\fbox{
\begin{minipage}{14.5cm}
	
\begin{center}
{ \sf Ticket-Issuer-Registration}
\end{center}

\begin{tabular}{lcl}
Ticket Issuer: $\mathcal{I}$ & & Central Authority: $\mathcal{CA}$\\
Selects $x_{i}\stackrel{R}{\leftarrow}\mathbb{Z}_{p}$, and computes  $Y_{I}=\xi^{x_{i}}$ \\
and $\tilde{Y}_{I}=\mathfrak{g}^{x_{i}}$\\
The secret-public key pair is $(x_{i},Y_{I},\tilde{Y}_{I})$. &$~~\xrightarrow{ID_{I},Y_{I},\tilde{Y}_{I}}$~~ &  Select $e_{i},r_{i}\stackrel{R}{\leftarrow}\mathbb{Z}_{p}$ and computes\\
Verifies: $e(\sigma_{I},{Y}_{A}\mathfrak{g}^{e_{i}})\stackrel{?}{=}e(gh^{r_{i}}{Y}_{I},\mathfrak{g})$& ~~$\xleftarrow{\sigma_{I},r_{i},e_{i}}$~~ &$\sigma_{I}=(gh^{r_{i}}{Y}_{I})^{\frac{1}{x_{a}+e_{i}}}$ \\
Keeps the credential as $Cred_{I}=(e_{i},r_{i},\sigma_{I})$ & &Stores $(ID_{I},Y_{I},\tilde{Y}_{I},(r_{i},e_{i},\sigma_{I}))$\\
\end{tabular}

\begin{center}
{\sf Ticket-Verifier-Registration}
\end{center}
\begin{tabular}{lcl}
Ticket-Verifier: $\mathcal{V}$ & & Central Authority: $\mathcal{CA}$\\
Selects $x_{v}\stackrel{R}{\leftarrow}\mathbb{Z}_{p}$ and computes  $Y_{V}=\xi^{x_{v}}$\\
The secret-public key pair is $(x_{v},Y_{V})$ &$~~\xrightarrow{ID_{V},Y_{V}}$~~ &   Selects $\lambda_{v},r_{v}\stackrel{R}{\leftarrow}\mathbb{Z}_{p}$ and computes\\

Verifies: $e(\sigma_{V},{Y}_{A}\mathfrak{g}^{e_{v}})\stackrel{?}{=}e(gh^{r_{v}}Y_{V},\mathfrak{g})$ & $\xleftarrow{\sigma_{V},r_{v},\lambda_{v}}$ &$\sigma_{V}=(gh^{r_{v}}Y_{V})^{\frac{1}{x_{a}+\lambda_{v}}}$\\
Keep the credential as $Cred_{V}=(\lambda_{v},r_{v},\sigma_{V})$& & Stores $(ID_{V},Y_{V},(r_{v},\lambda_{v},\sigma_{V}))$\\
\end{tabular}

\begin{center}
{\sf User-Registration}
\end{center}
\begin{tabular}{lcl}
User: $\mathcal{U}$ & & Central Authority: $\mathcal{CA}$\\
Selects $x_{u}\stackrel{R}{\leftarrow}\mathbb{Z}_{p}$, and computes $Y_{U}=\xi^{x_{u}}$ &&\\
This secret-public key pair is $(x_{u},Y_{U})$ & $~~\xrightarrow{ID_{U},Y_{U}}$ & Select $e_{u},r_{u}\stackrel{R}{\leftarrow}\mathbb{Z}_{p}$ and computes\\

Verifies: $e(\sigma_{U},Y_{A}\mathfrak{g}^{e_{u}})\stackrel{?}{=}e(gh^{r_{u}}{Y}_{U},\mathfrak{g})$& ~~$\xleftarrow{\sigma_{U},e_{u},r_{u}}$~~ & $\sigma_{U}=(gh^{r_{u}}{Y}_{U})^{\frac{1}{x_{a}+e_{u}}}$\\
Keep the credential as $Cred_{U}=(e_{u},r_{u},\sigma_{U})$.& & Stores $(ID_{U},Y_{U},(r_{u},e_{u}\sigma_{U}))$\\
\end{tabular}

\begin{center}
{\sf Central-Verifier-Registration}
\end{center}
\begin{tabular}{lcl}
Central Verifier: $\mathcal{CV}$ & & Central Authority: $\mathcal{CA}$\\
Selects $x_{cv}\stackrel{R}{\leftarrow}\mathbb{Z}_{p}$, and computes $Y_{CV}=\xi^{x_{cv}}$ & &\\
The secret-public key pair is $(x_{cv},Y_{CV})$ & $~~\xrightarrow{ID_{CV},Y_{CV}}$~~  & Select $\lambda_{cv},r_{cv}\stackrel{R}{\leftarrow}\mathbb{Z}_{p}$ and computes\\
Verifies: $e(\sigma_{cv},Y_{A}\mathfrak{g}^{\lambda_{cv}})\stackrel{?}{=}e(gh^{r_{cv}}{Y}_{CV},\mathfrak{g})$ & ~~$\xleftarrow{\sigma_{CV},\lambda_{cv},r_{cv}}$~~ & $\sigma_{CV}=(gh^{r_{cv}}{Y}_{CV})^{\frac{1}{x_{a}+\lambda_{cv}}}$\\
Keep the credential as $Cred_{CV}=(\lambda_{cv},r_{cv},\sigma_{CV})$ & & Stores $(ID_{CV},Y_{CV},(r_{cv},\lambda_{cv},\sigma_{CV}))$\\
\end{tabular}
\end{minipage}
}\caption{Registration Algorithm}\label{fig.reg}
\end{figure}

Fig. \ref{fig.reg} depicts the registration processes. When registering 
with the $\mathcal{CA}$, $\mathcal{I}$, $\mathcal{V}$, $\mathcal{U}$ 
and $\mathcal{CV}$ use the $PP$ and generate their own secret-public 
key pairs. They then send their identities and associated public keys 
to $\mathcal{CA}$ which, after receiving a registration request from an 
entity, uses $MSK$ to generate the corresponding credential for them.  
Note that only the ticket issuer has two public keys, $Y_I$ and 
$\tilde{Y}_I$. The first one is used to sign the tickets while the 
second one is used to validate the ticket.

\subsection{Ticket Issuing}

\begin{figure}[!h]
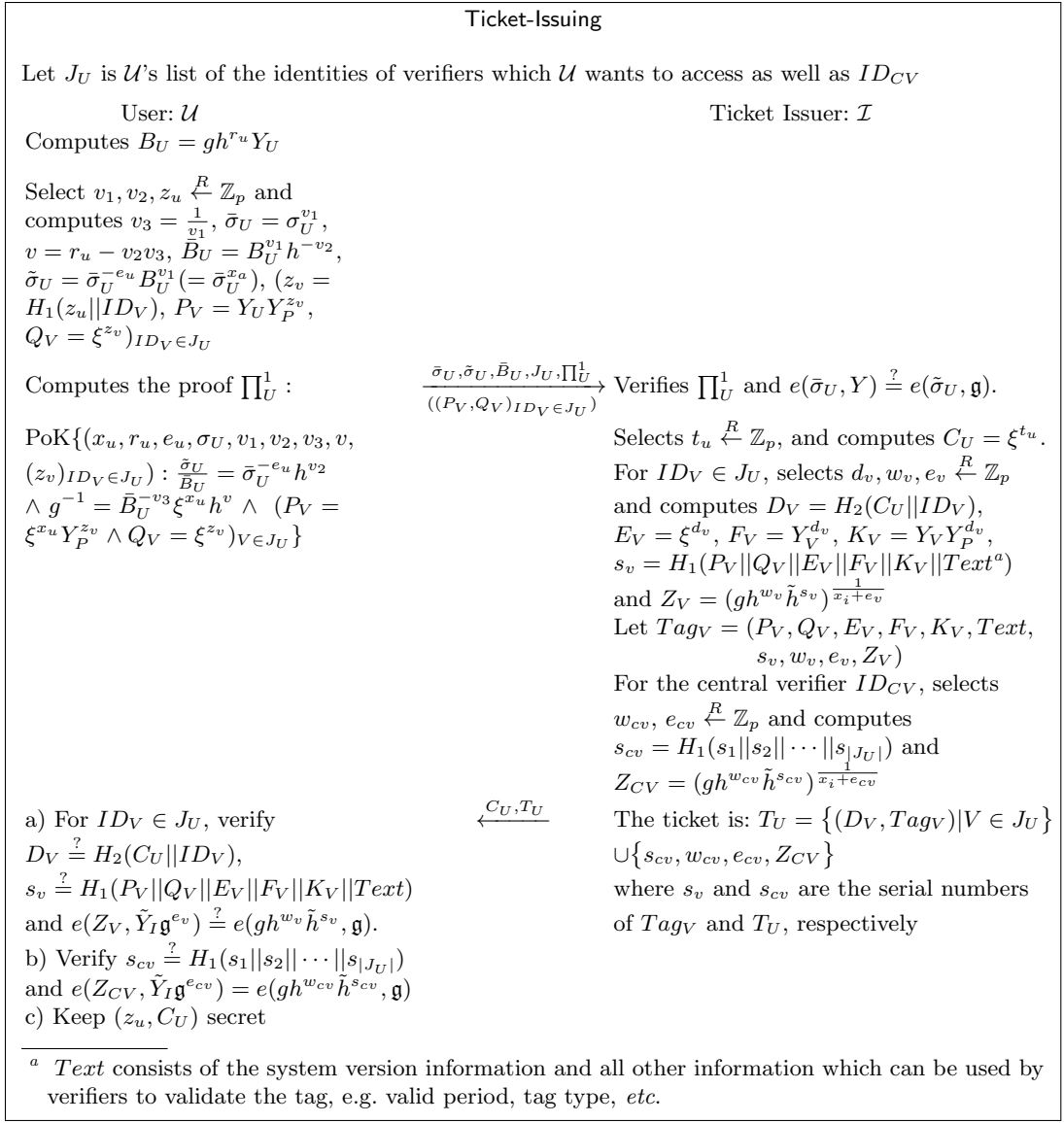

\centering
\fbox{
\begin{minipage}{14cm}
\begin{center}
{\sf Ticket-Issuing}
\end{center}

Let $J_{U}$ is $\mathcal{U}$'s list of the identities of verifiers 
which $\mathcal{U}$ wants to access as well as $ID_{CV}$ 
 \medskip
 
\begin{tabular}{lcl} {\hspace{1.2cm} User: $\mathcal{U}$} & & 
\hspace{1.2cm} Ticket Issuer: $\mathcal{I}$\\ \medskip

Computes $B_{U}=gh^{r_{u}}{Y}_{U}$ & &\\ 

Select $v_{1},v_{2},z_{u}\stackrel{R}{\leftarrow}\mathbb{Z}_{p}$ and \\ 
computes $v_{3}=\frac{1}{v_{1}}$, 
$\bar{\sigma}_{U}=\sigma_{U}^{v_{1}}$, \\ $v=r_{u}-v_{2}v_{3}$, 
$\bar{B}_{U}=B_{U}^{v_{1}}h^{-v_{2}}$,\\ 
$\tilde{\sigma}_{U}=\bar{\sigma}_{U}^{-e_{u}}B_{U}^{v_{1}}(= 
\bar{\sigma}_{U}^{x_{a}})$, $(z_{v}=$ \\
  $H_{1}(z_{u}||{ID_V}),$ $P_{V}=Y_{U}{Y}_{P}^{z_{v}},$\\
   $ Q_{V}=\xi^{z_{v}})_{{ID_V}\in J_{U}}$
\vspace{0.1cm} \\

Computes the proof $\prod_{U}^{1}:$ & 
$\xrightarrow[((P_{V},Q_{V})_{ID_{V}\in 
J_{U}})]{\bar{\sigma}_{U},\tilde{\sigma}_{U},\bar{B}_{U},J_{U}, 
\prod_{U}^{1}}$& Verifies $\prod_{U}^{1}$ and 
$e(\bar{\sigma}_{U},Y)\stackrel{?}{=} 
e(\tilde{\sigma}_{U},\mathfrak{g})$.\\

$\mbox{PoK}\large\{(x_{u},r_{u},e_{u},\sigma_{U},v_{1},v_{2},v_{3},v,$ 
& & Selects $t_{u}\stackrel{R}{\leftarrow}\mathbb{Z}_{p}$, and computes 
$C_{U}=\xi^{t_{u}}$.\\ $(z_{v})_{ID_{V}\in J_{U}}): 
\frac{\tilde{\sigma}_{U}}{\bar{B}_{U}}=\bar{\sigma}_{U}^{-e_{u}}h^{v_{2}}$ 
& & For $ID_{V}\in J_{U}$, selects 
$d_{v},w_{v},e_{v}\stackrel{R}{\leftarrow}\mathbb{Z}_{p}$\\ $ \wedge~ 
g^{-1}= \bar{B}_{U}^{-v_{3}}\xi^{x_{u}}h^{v}$ $\wedge~$ $(P_{V}= $ & & 
and   computes  $D_{V}=H_{2}(C_{U}||{ID_V})$, \\ 
$\xi^{x_{u}}{Y}_{P}^{z_{v}} \wedge Q_{V}=\xi^{z_{v}})_{{V}\in 
J_{U}}\large\}$ & &  $E_{V}=\xi^{d_{v}}$,   $F_{V}=Y_{V}^{d_{v}}$, 
$K_{V}=Y_{V}{Y}_{P}^{d_{v}}$, \\ 

& &$s_{v}=H_{1}(P_{V}||Q_{V}||E_{V}||F_{V}||K_{V}||Text$\footnote{ 
$Text$ consists of the system version information and all other 
information which can be used by verifiers to validate the tag, e.g. 
valid period, tag type, {\em etc}.}$)$  \\

& & and $Z_{V}=(gh^{w_{v}}\tilde{h}^{s_{v}})^{\frac{1}{x_{i}+e_{v}}}$\\
& & Let $Tag_V=(P_{V},Q_{V},E_{V},F_{V},K_{V},Text,$\\
& & ~~~~~~~~~~~~~~~~~~$s_{v},w_{v},e_{v},Z_{V})$\\
& & For the central verifier $ID_{CV}$, selects\\
& & $w_{cv}$, $e_{cv}\stackrel{R}{\leftarrow}\mathbb{Z}_{p}$ and computes\\
& & $s_{cv}=H_{1}(s_{1}||s_{2}||\cdots||s_{|J_{U}|})$ and \\
& & $Z_{CV}=(gh^{w_{cv}}\tilde{h}^{s_{cv}})^{\frac{1}{x_i+e_{cv}}}$\\

a) For $ID_{V}\in J_{U}$, verify & $\xleftarrow{C_{U},T_{U}}$ & The 
ticket is: $T_{U}=\big\{(D_{V},Tag_{V})|{V}\in J_{U}\big\}$\\ 
$D_{V}\stackrel{?}{=}H_{2}(C_{U}||{ID_V})$,& &  $\cup\big\{s_{cv},w_{cv},e_{cv},Z_{CV}\big\}$ \\ 
$s_{v}\stackrel{?}{=}H_{1}(P_{V}||Q_{V}||E_{V}||F_{V}||K_{V}||Text)$ &&where $s_{v}$ and $s_{cv}$ are the serial numbers\\
and $e(Z_{V},\tilde{Y}_{I}\mathfrak{g}^{e_{v}})\stackrel{?}{=} 
e(gh^{w_{v}}\tilde{h}^{s_{v}},\mathfrak{g})$. & & of 
$Tag_{V}$ and $T_U$, respectively\\ 
b) Verify $s_{cv}\stackrel{?}{=}H_{1}(s_{1}||s_{2}||\cdots||s_{|J_{U}|})$ & & \\ 
and $e(Z_{CV},\tilde{Y}_{I}\mathfrak{g}^{e_{cv}})= 
e(gh^{w_{cv}}\tilde{h}^{s_{cv}},\mathfrak{g})$& &\\ 

c) Keep $(z_{u},C_{U})$ secret\\ 

\end{tabular}
\end{minipage} }
\caption{Ticket Issuing Algorithm}\label{fig.t-i} 
\end{figure}

During the ticket issuing process (shown in Fig. \ref{fig.t-i}), the user
$\mathcal{U}$ defines $J_{U}$ to be  the set containing the identities of the
ticket verifiers whose services she wants to access as well as the identity of
the central verifier. In order to request a ticket from $\mathcal{I}$,
$\mathcal{U}$ creates pseudonyms, $P_V$ and $Q_V$, for each ${ID_V}\in J_{U}$ by
using her secret key to protect the anonymity of the verifiers. She also
produces a proof of knowledge of her credentials and submits this proof together
with the set $J_U$ and the pseudonyms to $\mathcal{I}$ to convince him that she
is a registered user and created the pseudonyms. Once $\mathcal{I}$ has received
this information and verified the proof of knowledge, he generates an
authentication tag $Tag_{V}$ for each ${ID_V}\in J_{U}$ as well as an overall
$Tag_{CV}$ for $\mathcal{CV}$ in case the ticket needs to be traced. Note that
these tags are constructed using the public keys of the respective verifiers and
thus can only be validated by the corresponding $\mathcal{V}$ or the central
verifier, $\mathcal{CV}$. The ticket is formed from these individual tags. Note
that each tag and the overall ticket are signed by the issuer using his private
key while the integrity of the tags and the overall ticket is assured using
hashes of their respective content. The ticket is sent back to $\mathcal{U}$ who
verifies the integrity of each tag and the overall ticket using the supplied
hash values as well as that each tag and the overall ticket have been signed by
the issuer.

\subsection{Tag Verification}
\label{subsec:TagVerify}

\begin{figure}[!h]
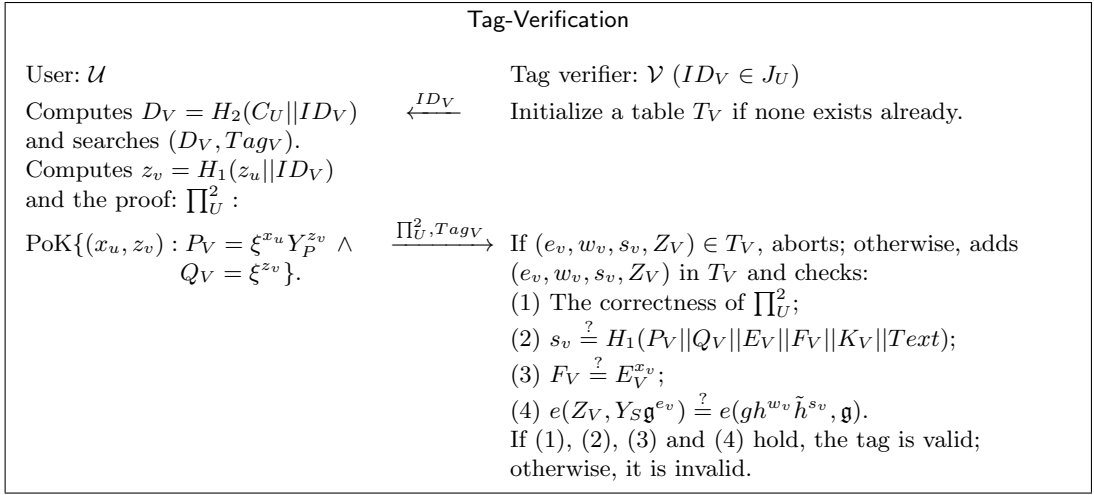

\centering
\fbox{
\begin{minipage}{14cm}
\begin{center}
{\sf Tag-Verification}
\end{center}
\begin{tabular}{lcl}
User: $\mathcal{U}$ && Tag verifier: $\mathcal{V}$ $(ID_{V}\in J_{U})$\\
Computes $D_{V}=H_{2}(C_{U}||{ID_V})$ & $\xleftarrow{ID_{V}}$ & Initialize a table $T_{V}$ if none exists already.\\
and searches $(D_{V},Tag_{V}).$\\
Computes $z_{v}=H_{1}(z_{u}||{ID_V})$  \\
and the proof: $\prod_{U}^{2}:$\\
$\mbox{PoK}\{(x_{u},z_{v}): P_{V}=\xi^{x_{u}}{Y}_{P}^{z_{v}}~\wedge$ &~~ $\xrightarrow{\prod_{U}^{2},Tag_{V}}$~~&  If  $(e_{v},w_{v},s_{v},Z_{V})\in T_{V}$, aborts; otherwise, adds\\
$~~~~~~~~~~~~~~~~~~~Q_{V}=\xi^{z_{v}}\}$. & & $(e_{v},w_{v},s_{v},Z_{V})$ in $T_{V}$ and checks:\\
& & (1) The correctness of $\prod_{U}^{2}$;\\
& & (2) $s_{v}\stackrel{?}{=}H_{1}(P_{V}||Q_{V}||E_{V}||F_{V}||K_{V}||Text)$;\\
& & (3) $F_{V}\stackrel{?}{=}E_{V}^{x_{v}}$;\\
& & (4)  $e(Z_{V},Y_{S}\mathfrak{g}^{e_{v}})\stackrel{?}{=}e(gh^{w_{v}}\tilde{h}^{s_{v}},\mathfrak{g})$.\\
& & If (1), (2), (3) and (4)  hold, the tag is valid; \\
& & otherwise, it is invalid. \\
\end{tabular}
\end{minipage}
}\caption{Tag Verification Algorithm}\label{fig.t-v}
\end{figure}

The tag verification process is shown in Fig. \ref{fig.t-v}. When the 
user $\mathcal{U}$ wants to access a service, the tag verifier 
$\mathcal{V}$ send his identity information to the user which 
$\mathcal{U}$  uses to look up the corresponding tag, $Tag_V$. In order 
to access the service, $\mathcal{U}$ must submit a proof of knowledge 
of her secret key alongside the relevant authentication tag $Tag_{V}$ 
to prevent users from sharing authentication tags. $\mathcal{V}$ checks 
his table of previously received tags to ensure that the tag has not 
already been used previously (double-spend detection), before verifying 
the user's proof of knowledge in Step 1. Step 2 checks the integrity of 
the tag using a hash function while Step 4 verifies that it has been 
issued by the ticket issuer, $\mathcal{I}$. Step 3 can only be verified 
by $\mathcal{V}$ as it requires the private key of the verifier. Only 
if $\mathcal{V}$ can complete all steps successfully, is the user 
granted access.


\subsection{Ticket Tracing}
\begin{figure}[!h]
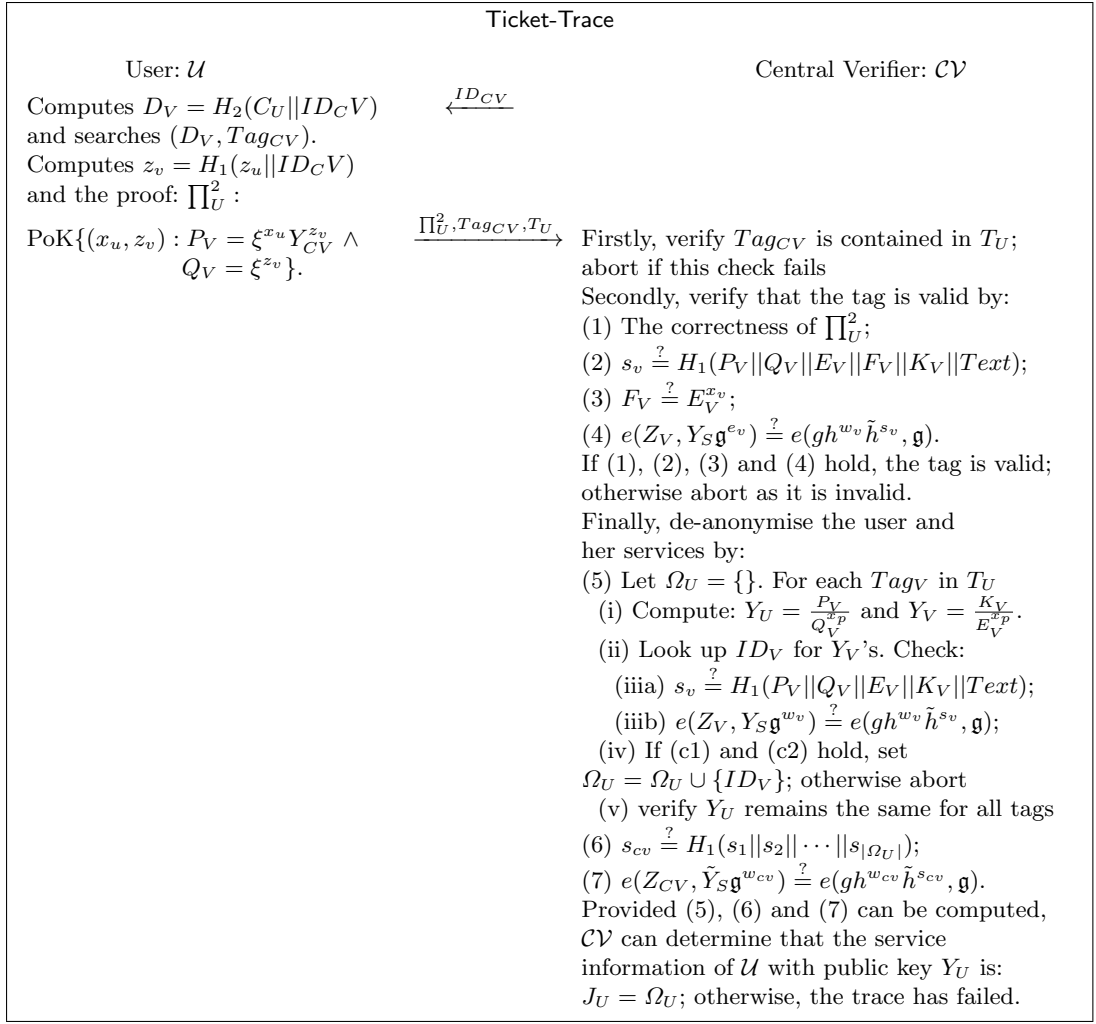

\centering
\fbox{
\begin{minipage}{14cm}
\begin{center}{\sf Ticket-Trace}
\end{center}
\begin{tabular}{lcl}
\hspace{1.2cm} User: $\mathcal{U}$ &\hspace{2.5cm} & \hspace{2.2cm} Central Verifier: $\mathcal{CV}$\\
Computes $D_{V}=H_{2}(C_{U}||{ID_CV})$ & $\xleftarrow{ID_{CV}}$ &\\
and searches $(D_{V},Tag_{CV}).$\\
Computes $z_{v}=H_{1}(z_{u}||{ID_CV})$  \\
and the proof: $\prod_{U}^{2}:$\\
$\mbox{PoK}\{(x_{u},z_{v}): P_{V}=\xi^{x_{u}}{Y}_{CV}^{z_{v}}~\wedge$ &~~ $\xrightarrow{\prod_{U}^{2},Tag_{CV}, T_U}$~~& Firstly, verify $Tag_{CV}$ is contained in $T_U$; \\
$~~~~~~~~~~~~~~~~~~~Q_{V}=\xi^{z_{v}}\}$. & & abort if this check fails\\
& & Secondly, verify that the tag is valid by:\\
& & (1) The correctness of $\prod_{U}^{2}$;\\
& & (2) $s_{v}\stackrel{?}{=}H_{1}(P_{V}||Q_{V}||E_{V}||F_{V}||K_{V}||Text)$;\\
& & (3) $F_{V}\stackrel{?}{=}E_{V}^{x_{v}}$;\\
& & (4)  
$e(Z_{V},Y_{S}\mathfrak{g}^{e_{v}})\stackrel{?}{=}e(gh^{w_{v}}\tilde{h}^{s_{v}},\mathfrak{g})$.\\
& & If (1), (2), (3) and (4)  hold, the tag is valid; \\
& & otherwise abort as it is invalid. \\
& & Finally, de-anonymise the user and\\
& & her services by:\\
 & &(5) Let $\Omega_{U}=\{\}$. For each $Tag_V$ in $T_U$\\
 & & ~~(i) Compute: $Y_{U}=\frac{P_{V}}{Q_{V}^{x_{p}}}$ and 
 $Y_{V}=\frac{K_{V}}{E_{V}^{x_{p}}}$.\\
 & & ~~(ii) Look up $ID_V$ for $Y_V$'s. Check: \\
 & & ~~~~(iiia) $s_{v}\stackrel{?}{=}H_{1}(P_{V}||Q_{V}||E_{V}||K_{V}||Text)$;\\
 & & ~~~~(iiib) $e(Z_{V},Y_{S}\mathfrak{g}^{w_{v}})\stackrel{?}{=} 
 e(gh^{w_{v}}\tilde{h}^{s_{v}},\mathfrak{g})$;\\
 & & ~~(iv) If (c1) and (c2) hold, set \\
 & & $\Omega_U=\Omega_U\cup\{ID_V\}$; otherwise abort\\
 & & ~~(v) verify $Y_U$ remains the same for all tags\\
 & & (6) $s_{cv}\stackrel{?}{=}H_{1}(s_{1}||s_{2}||\cdots||s_{|\Omega_{U}|})$;\\
 & & (7) 
 $e(Z_{CV},\tilde{Y}_{S}\mathfrak{g}^{w_{cv}})\stackrel{?}{=}e(gh^{w_{cv}}\tilde{h}^{s_{cv}},\mathfrak{g})$.\\
 & & Provided (5), (6) and (7) can be computed, \\
 & & $\mathcal{CV}$ can determine that the service\\
 & & information of $\mathcal{U}$ with public key $Y_{U}$ is:\\
 & & $J_{U}=\Omega_U$; otherwise, the trace has failed.\\
\end{tabular}
\end{minipage}
}\caption{Ticket Trace Algorithm}\label{fig.t-t}
\end{figure}

Lastly, in the case that a user $\mathcal{U}$'s whole service information $J_U$
needs to be traced, the central verifier, $\mathcal{CV}$, sends its identity to
$\mathcal{U}$ who is then required to submit the information needed by the
Ticket Validation algorithm as well as her overall ticket. Note that, provided a
single tag is known, the whole ticket information could also be obtained
directly from the issuer, $\mathcal{I}$, in case the user is not co-operating.

On receipt of this information, the central verifier first validates that the
submitted tag $Tag_{CV}$  passes the standard verification process (see
Section\ref{subsec:TagVerify}) as the central verifier's $ID_{CV}$ is always
included in $J_U$. As discussed previously, this steps ensures that
$\mathcal{U}$  is a valid user and that the tag belongs to her. Once this steps
has passed, the central verifier can then validate the integrity of the ticket
and that the previously presented authentication tag is indeed part of the
ticket which establishes that the ticket does indeed belong to user who
presented it. Using his private key, the central verifier can now compute the
user $\mathcal{U}$'s public key as well as the public keys of all the verifiers
contained within the authentication tags and thus determine the user's identity
and her service information $J_{U}$.

\section{Security Analysis} \label{sec:security}

In this section we present the theorems which establish the security of our
scheme. 

\begin{theorem}[\bf Unlinkability]%
An anonymous Single-Sign-On for $n$ designated services with traceability scheme
in Fig. \ref{fig.setup}, Fig. \ref{fig.reg}, Fig. \ref{fig.t-i}, Fig.
\ref{fig.t-v} and Fig.  \ref{fig.t-t} is 
$\boldsymbol{(\rho_{1},\rho_{2},\rho_{3},\epsilon'(\ell))}${\bf-selectively 
unlinkable} if 
the DDH
assumption holds on the bilinear group
$(e,p,\mathbb{G}_{1},\mathbb{G}_{2},\mathbb{G}_{\tau})$ with the advantage at
most  $\epsilon(\ell)$, and $H_{1}, H_{2}$ are  secure cryptographic hash
functions, where $\varrho_{1}$ is the total number of verifiers selected by
$\mathcal{A}$ to query tickets,  $\varrho_{2}$ is the number of ticket
validation queries, $\varrho_{3}$ is the number of ticket trace queries,
$\epsilon(\ell)=\frac{\epsilon'(\ell)}{2}$.

\label{theorem:unlink}
\end{theorem}

The proof of Theorem \ref{theorem:unlink} follows 
from the unlinkability game in Appendix \ref{unlink-game} and is formally 
proved in Appendix~\ref{app:unlink}.{}

\begin{theorem}[\bf Unforgeability]%
An anonymous Single-Sign-On for n designated services with
traceability scheme  in Fig. 
\ref{fig.setup}, Fig. \ref{fig.reg}, Fig. \ref{fig.t-i}, Fig. \ref{fig.t-v} and 
Fig.  \ref{fig.t-t} is  
$\boldsymbol{(\varrho,\epsilon'(\ell))}${\bf-unforgeable} if the 
JOC-version-$q$-SDH assumption holds on the bilinear group 
$(e,p,\mathbb{G}_{1},\mathbb{G}_{2},\mathbb{G}_{\tau})$ with the advantage at 
most  $\epsilon(\ell)$,  and $H_{1},H_{2}$ are secure cryptographic hash 
functions, where $\varrho$ is the total number of verifiers selected by 
$\mathcal{A}$ to query tickets,  $\varrho\leq q$, 
$\epsilon(\ell)=(\frac{p-q}{p}+\frac{1}{p}+\frac{p-1}{p^{3}})\epsilon'(\ell)$.
\label{theorem:unforge}
\end{theorem}

Theorem \ref{theorem:unforge} is demonstrated by the
unforgeability game in Appendix \ref{unforge-game} and it is formally proved in
Appendix~\ref{app:unforge}.{}

\begin{theorem}[\bf Traceability]%
An anonymous Single-Sign-On for n designated services with
traceability scheme in Fig.
\ref{fig.setup}, Fig. \ref{fig.reg}, Fig. \ref{fig.t-i}, Fig. \ref{fig.t-v} and
Fig.  \ref{fig.t-t} is $\boldsymbol{(\rho,\epsilon(\ell))}${\bf-traceable} if 
the 
$q$-SDH
assumption holds on the bilinear group
$(e,p,\mathbb{G}_{1},\mathbb{G}_{2},\mathbb{G}_{\tau})$ with the advantage at
most  $\epsilon_{1}(\ell)$, the DL assumption holds on the group
$\mathbb{G}_{1}$ with the advantage at most $\epsilon_{2}(\ell)$, and $H_{1},
H_{2}$ are  secure cryptographic hash functions, where
$\epsilon(\ell)=max\left\{\frac{\epsilon_{1}(\ell)}{2}(\frac{p-q}{p}+
\frac{1}{p}+\frac{p-1}{p^{3}}),\frac{\epsilon_{2}(\ell)}{2}\right\}$, $\varrho$
is the total number of ticket issuing queries made by $\mathcal{A}$ and
$\varrho<q$.%
\label{theorem:trace}
\end{theorem}

Theorem \ref{theorem:trace} follows from the 
traceability game in Appendix~\ref{trace-game} and its formal proof is given in 
Appendix \ref{app:trace}.{}

\section{Benchmarking results} 
\label{sec:benchmarks}

Our proposed scheme has been implemented (\cite{githubrepo}) on a Dell Inspiron
Latitude E5270 laptop with an Intel Core i7-6600U CPU, 1TB SSD and 16GB of RAM
running Fedora 27.

The implementation makes use of bilinear pairings using elliptic curves as well
as other cryptographic primitives. As such the implementation of the scheme
relies on the JPBC library~(\cite{jpbc}) for the bilinear pairings and uses the
cryptographic functions provided by bouncycastle~(\cite{bouncycastle}). 

\subsection{Timings} 

Table~\ref{tab:Benchmarks} shows the results of the computational time spent in
the various phases of our proposed scheme which required more complex
computations (\ie some form of verification or generation of proofs) . Note that
the Java based implementation of the JPBC API~(\cite{jpbc}) was used throughout.

The bilinear mapping used in the protocol implementations was a 
Type~F elliptic curve
%
%
%
where $G$ is the group of points of the elliptic curve and
$|G|=p$ is its prime order whose binary representation requires $r$-bits.
 
\inserttabd{tab:Benchmarks}{|l|c|c|c|c|c|}
{Benchmark results {(in ms)}} 
{
Protocol phase &Entity&\multicolumn{4}{|c|}{(r,V), r=\#bits; 
V=\#verifiers } 
\\
&&$(160,2)$ &$(160,3)$ & $(320,2)$ &$(320,3)$\\
\hline
\multicolumn{6}{|c|}{System Initialisation -  Central Authority 
($\mathcal{CA}$)}\\
\hline
initialise the system &CA 
&\multicolumn{2}{|c|}{1398}&\multicolumn{2}{|c|}{3385}\\
\hline
\multicolumn{6}{|c|}{Registration - Issuer ($\mathcal{I}$)}\\
\hline
generate I credentials 
&CA&\multicolumn{2}{|c|}{12}&\multicolumn{2}{|c|}{45}\\
\hline
verify I credentials &I&\multicolumn{2}{|c|}{641}&\multicolumn{2}{|c|}{979}\\
\hline
\multicolumn{6}{|c|}{Registration - User ($\mathcal{U}$)}\\
\hline
generate user credentials 
&CA&\multicolumn{2}{|c|}{12}&\multicolumn{2}{|c|}{20}\\
\hline
verify user credentials  
&User&\multicolumn{2}{|c|}{301}&\multicolumn{2}{|c|}{498}\\
\hline
\multicolumn{6}{|c|}{Registration - Central Verifier ($\mathcal{CV}$)}\\
\hline
generate CV credentials &CA&\multicolumn{2}{|c|}{9}&\multicolumn{2}{|c|}{23}\\
\hline
verify CV credentials &CV&\multicolumn{2}{|c|}{269}&\multicolumn{2}{|c|}{497}\\
\hline
\multicolumn{6}{|c|}{Registration - Verifier ($\mathcal{V}$)}\\
\hline
generate V credentials &CA&\multicolumn{2}{|c|}{10}&\multicolumn{2}{|c|}{23}\\
\hline
verify V credentials & V &\multicolumn{2}{|c|}{290}&\multicolumn{2}{|c|}{623}\\
\hline
\multicolumn{6}{|c|}{Issuing phase}\\
\hline
generate $\Pi_U^1$ \& ticket request &User &93&101&280&309 \\
\hline
verify $\Pi_U^1$, generate ticket & Issuer &481&515&916&1044\\
\hline
verify ticket  &User &764&960&1960&2567\\
\hline
\multicolumn{6}{|c|}{Tag Verification - Verifier ($\mathcal{V}$)}\\
\hline
retrieve $Tag_V$ \& generate $\Pi_U^2$ &User 
&\multicolumn{2}{|c|}{13}&\multicolumn{2}{|c|}{34}\\
\hline
verify $\Pi_U^2$ \& $Tag_V$ & V  
&\multicolumn{2}{|c|}{225}&\multicolumn{2}{|c|}{575}\\
\hline
\multicolumn{6}{|c|}{Ticket Tracing - Central Verifier ($\mathcal{CV}$)}\\
\hline
retrieve ticket $T_U$ \& $Tag_{CV}$; generate $\Pi_U^2$ &User &8&9&33&37\\
\hline
verify $\Pi_U^2$, $Tag_{CV}$; trace $T_U$ & CV &983&1146&2575&3182\\
}

The creation of credentials by the CA for the issuer, user and the (central)
verifiers during the registration phase of the protocol is on average around
$12$ms for $r=160$ bits and $30$ms for $r=320$ bits while the verification of
those credentials by the various parties takes about $300$ms and $650$ms for
$160$ bits and $320$ bits respectively.

It can be seen from Table~\ref{tab:Benchmarks} that the current implementation
of the our scheme is reasonably fast for elliptic curves where $r=160$ (\eg
$\approx 1.5$s and $\approx 250$ms for ticket issuing and verification
respectively) and still acceptable for $r=320$ bits ($\approx 4$s and $\approx
600$ms for the same steps).

Moreover, it should be possible to improve the performance of the code by
switching from the current Java-based version to using a Java-wrapper to the
C-based implementation of the pbc libraries~(\cite{cpbc}), instead.

\section{Conclusion and Future Work}\label{sec:conclusion} %
Previous Anonymous Single-Sign-On schemes usually protect the user's identity
from other verifiers but not always the issuer nor the verifier to whom the user
needs to authenticate. However, previously, the identity of these verifiers has
not been considered extensively and neither has the need to ensure that only a
designated verifier can validate a given access request. In this paper we
proposed an Anonymous Single-Sign-On scheme which enables users and verifiers to
remain anonymous throughout while protecting the system from misbehaving users
through a central verifier who can, if required, trace the identities of a user
and her associated verifiers.  Moreover, we provided a formal security model and proofs for the security
properties of our scheme as well as an implementation demonstrating the
feasibility of deployment. %

In our scheme, a user can currently only authenticate to a verifier once as
there is only one authentication tag for each verifier in a user's ticket. If
the user needs to authenticate herself to a verifier, $\mathcal{V}$, multiple
times, she must request additional tickets with the required authentication tag
for $\mathcal{V}$ from the issuer. Our scheme could alternatively be amended to
allow multiple authentication tags per verifier in each ticket.  In this case the
scheme's security model and proofs would need to be amended to support this.

Anonymous Single-Sign-On was the main motivational use case for our scheme, but
there are other scenarios to which the could be applied, \eg  the purchase of
tickets for tourist attractions, where being able to issue a ticket through an
Android implementation would be appropriate. Initial results however demonstrate
that the timings on an Android client are significantly slower, for example
ticket validation can take $\approx 200$ times longer than on the laptop. Future
work will focus on improving the scheme's performance further (especially on the
Android platform) by moving from a pure Java-based implementation to a C-based
version.

Lastly, extending our scheme with an option for users to enable the controlled
release of personal information to a given verifier, \eg by letting a user
control which verifier is allowed to deanonymise her authentication tag, is
another area of future research.

\medskip%
\noindent{\sf\bf Acknowledgement:} This work has been supported by the EPSRC
Project DICE: ``Data to Improve the Customer Experience'', EP/N028295/1.\\ %
The authors would also like to thank Dr Fran\c{c}ois Dupressoir for his 
valuable feedback on an early draft of this paper.

\appendix
\section{Formal Definition}
\label{app:formal_def_alg}

The definition of our scheme is formalised by the following five algorithms:
\begin{enumerate}
\item{\sf Setup}
$(1^{\ell})\rightarrow \left(MSK,PP\right).$ $\mathcal{CA}$ inputs a security
parameter $1^{\ell}$, and outputs the master secret key $MSK$ and the public
parameters $PP$. \medskip

\item{\sf Registration}: This algorithm consists of the following four sub-algorithms:
\medskip
\begin{itemize}
\item{\sf Ticket-Issuer-Registration} 
$(\mathcal{I}(ID_{I},SK_{I},PK_{I},PP)\leftrightarrow
\mathcal{CA}(MSK,PK_{I},PP))\rightarrow(\sigma_{I},(ID_{I},$ $PK_{I})).$ This is
an interactive algorithm executed between $\mathcal{CA}$ and $\mathcal{I}$.
$\mathcal{I}$ runs the secret-public key pair generation algorithm
$\mathcal{KG}(1^{\ell})\rightarrow(SK_{I},PK_{I})$, inputs its identity
$ID_{I}$, secret-public key pair $(SK_{I},PK_{I})$ and the public parameters
$PP$, and outputs a credential $\sigma_{I}$. $\mathcal{CA}$ inputs the master
secret key $MSK$, $\mathcal{I}$'s public key $PK_{I}$ and the public parameters
$PP$, and outputs the identity $ID_{I}$ and the public key $PK_{I}$. \medskip

\item{\sf Ticket-Verifier-Registration} 
$(\mathcal{V}(ID_{V},SK_{V},PK_{V},PP)\leftrightarrow
\mathcal{CA}(MSK,PK_{V},PP))\rightarrow(\sigma_{V},$ \\$(ID_{V}, PK_{V}))$. This
is an interactive algorithm executed between $\mathcal{CA}$  and $\mathcal{V}$.
$\mathcal{V}$ runs $\mathcal{KG}(1^{\ell})\rightarrow(SK_{V},$ $PK_{V})$, inputs
its identity $ID_{V}$, secret-public key pair $(SK_{V},PK_{V})$ and the public
parameters $PP$, and outputs a credential $\sigma_{V}$. $\mathcal{CA}$ inputs
the master secret key $MSK$, $\mathcal{V}$'s public key $PK_{V}$ and the public
parameters $PP$, and outputs the identity $ID_{V}$ and the public key $PK_{V}$.
\medskip

\item{\sf User-Registration} 
$(\mathcal{U}(ID_{U},SK_{U},PK_{U},PP)\leftrightarrow
\mathcal{CA}(MSK,PK_{U},PP))\rightarrow(\sigma_{U},(ID_{U},$ $PK_{U})).$ This is
an interactive algorithm executed between $\mathcal{CA}$  and $\mathcal{U}$.
$\mathcal{U}$ runs $\mathcal{KG}(1^{\ell})\rightarrow(SK_{U},PK_{U})$, inputs
its identity $ID_{U}$, secret-public key pair $(SK_{U},PK_{U})$ and the public
parameters $PP$, and outputs a credential $\sigma_{U}$. $\mathcal{CA}$ inputs
the master secret key $MSK$, $\mathcal{U}$'s public key $PK_{U}$ and the public
parameters $PP$, and outputs the identity $ID_{U}$ and the public key $PK_{U}$.
\medskip

\item{\sf Central-Verifier-Registration} 
$(\mathcal{P}(ID_{CV},SK_{CV},PK_{CV},PP)\leftrightarrow
\mathcal{CA}(MSK,PK_{CV},PP))\rightarrow(\sigma_{CV},$ $(ID_{CV},PK_{CV})).$
This is an interactive algorithm executed between $\mathcal{CA}$  and
$\mathcal{CV}$. $\mathcal{CV}$ runs
$\mathcal{KG}(1^{\ell})\rightarrow(SK_{CV},PK_{CV})$, inputs its identity
$ID_{CV}$, secret-public key pair $(SK_{CV},$ $PK_{CV})$ and the public parameters
$PP$, and outputs a credential $\sigma_{CV}$. $\mathcal{CA}$ inputs the master
secret key $MSK$, $\mathcal{CV}$'s public key $PK_{CV}$ and the public
parameters $PP$, and outputs the identity $ID_{CV}$ and the public key
$PK_{CV}$.
\end{itemize}
\medskip

\item{\sf Ticket-Issuing}
$\left(\mathcal{U}(SK_{U},PK_{U},J_{U},\sigma_{U},PP)\leftrightarrow
\mathcal{I}(SK_{I},PK_{I},PP)\right)\rightarrow (T_{U},J_{U}).$ This is an
interactive algorithm executed between  $\mathcal{U}$ and  $\mathcal{I}$.
$\mathcal{U}$ takes as input his secret-public key pair $(SK_{U},PK_{U})$, his
service information $J_{U}$ consisting of the identities of ticket verifiers, 
his credential $\sigma_{U}$  and the public parameters $PP$, and outputs a
ticket  $T_{U}=Tag_{CV}\cup\{Tag_{V}|V\in J_{U}\}$ where the
authentication tags $Tag_{V}$ and $Tag_{CV}$ can only be validated by the
verifier $\mathcal{V}$ with $V\in J_{U}$ and the central verifier
$\mathcal{CV}$, respectively. $\mathcal{I}$ takes as input his secret-public key
pair $(SK_{I},PK_{I})$ and the public parameters $PP$, and outputs the service
information $J_{U}$. \medskip

\item{\sf Ticket-Validation}
$(\mathcal{U}(SK_{U},PK_{U},Tag_{V},PP)\leftrightarrow
\mathcal{V}((SK_{V},PK_{V}),PK_{I},PP))\rightarrow(\bot,(1,Tag_{V})$
$/(0,Tag_{V}))$. This is an interactive algorithm executed between $\mathcal{U}$
and $\mathcal{V}$ with $V\in J_{U}$. $\mathcal{U}$ takes as input his
secret-public key pair $(SK_{U},PK_{U})$,  the authentication tag $Tag_{V}$
and the public parameters $PP$, and outputs $\bot$. $\mathcal{V}$ takes input
his secret-public  key  pair $(SK_{V},PK_{V})$, $\mathcal{I}$'s public key
$PK_{S}$ and the public parameters $PP$, and outputs  $(1,Tag_{V})$ if ${V} \in
J_{U}$ and the authentication tag $Tag_{V}$ is valid; otherwise, it outputs 
$(0, Tag_{V})$ to indicate an invalid tag. \medskip 

\item{\sf Ticket-Trace}{
$(SK_{CV},PK_{CV},Tag_{CV},T_{U},PP))\rightarrow (\mathcal{U},J_{U})$. 
$\mathcal{CV}$ takes as inputs his secret-public key pair $(SK_{CV},PK_{CV})$,
the authentication token $Tag_{CV}$, the ticket $T_{U}$ and the public parameters $PP$, and outputs
$\mathcal{U}$ and $\mathcal{U}$'s  whole service set $J_{U}$.}

\end{enumerate}
\begin{definition}
An anonymous single-sign-on for $n$ designated services with
traceability is correct if 

\begin{equation*}
\Pr\left[\begin{array}{l|l}
                         & {\sf Setup}(1^{\ell})\rightarrow \left(MSK,PP\right); \\
                         &  {\mbox{\sf Ticket-Issuer-Registration}} (\mathcal{S}(ID_{I},SK_{I},PK_{I},PP)\leftrightarrow \\
                         &        \mathcal{CA}(MSK, PK_{I},PP))\rightarrow (\sigma_{I},(ID_{I},PK_{I}));\\
  &  {\mbox{\sf Ticket-Verifier-Registration}} (\mathcal{V}(ID_{V},SK_{V},PK_{V},PP)\leftrightarrow   \\
{\mbox{\sf Ticket-Validation}} (\mathcal{U}(SK_{U},  & \mathcal{CA}(MSK, PK_{V},PP))\rightarrow (\sigma_{V},(ID_{V},PK_{V}));\\
PK_{U},Tag_{V},  PP)\leftrightarrow                   &  {\mbox{\sf User-Registration}} (\mathcal{U}(ID_{U},SK_{U},PK_{U},PP)\leftrightarrow \mathcal{CA}(MSK,\\
 \mathcal{V}((SK_{V},PK_{V}), PK_{I},          & PK_{U},PP))\rightarrow (\sigma_{U},(ID_{U},PK_{U}));\\
 PP)) \rightarrow(\bot,  (1,Tag_{V}) )               & {\mbox{\sf Central-Verifier-Registration}} (\mathcal{S}(ID_{CV},SK_{CV},PK_{CV},PP) \\
                   &\leftrightarrow\mathcal{CA}(MSK,PK_{CV},PP))\rightarrow (\sigma_{P},(ID_{CV},PK_{CV}));\\
                   & {\mbox{\sf Ticket-Issuing}}(\mathcal{U}(SK_{U},PK_{U},J_{U},\sigma_{U},PP)\leftrightarrow \mathcal{I}(SK_{I},\\
                   & PK_{I},PP))\rightarrow  (T_{U},J_{U});\\
                   & {V}\in J_{U}
\end{array}
\right]=1
\end{equation*}

and 

\begin{equation*}
\Pr\left[\begin{array}{l|l}
                         & {\sf Setup}(1^{\ell})\rightarrow \left(MSK,PP\right); \\
                         &  {\mbox{\sf Ticket-Seller-Registration}} (\mathcal{I}(ID_{I},SK_{I},PK_{I},PP)\leftrightarrow \\
                        &        \mathcal{CA}(MSK, PK_{S},PP))\rightarrow (\sigma_{I},(ID_{I},PK_{I}));\\
     & {\mbox{\sf Ticket-Verifier-Registration}} (\mathcal{V}(ID_{V},SK_{V},PK_{V},PP)\leftrightarrow\\
  {\mbox{\sf Ticket-Trace}}(SK_{CV},PK_{CV}, &  \mathcal{CA}(MSK,PK_{V},PP))\rightarrow (\sigma_{V},(ID_{V},PK_{V}));\\
  Tag_{CV}, T_{U}, PP) \rightarrow(\mathcal{U},J_{U})  & {\mbox{\sf User-Registration}} (\mathcal{U}(ID_{U},SK_{U},PK_{U},PP)\leftrightarrow \mathcal{CA}(MSK, \\
     & PK_{U},PP))\rightarrow (\sigma_{U},(ID_{U},PK_{U}));\\
& {\mbox{\sf Central-Verifier-Registration}} (\mathcal{S}(ID_{CV},SK_{CV},PK_{CV},PP)\\
&\leftrightarrow \mathcal{CA}(MSK,PK_{CV},PP))\rightarrow (\sigma_{P},(ID_{CV},PK_{CV}));\\
              & {\mbox{\sf Ticket-Issuing}}(\mathcal{U}(SK_{U},PK_{U},J_{U},\sigma_{U},PP)\leftrightarrow \mathcal{I} (SK_{I},\\
          &PK_{I},PP))\rightarrow  (T_{U},J_{U}).
\end{array}
\right]=1.
\end{equation*}
\end{definition}

\section{Security Model}%
\label{sec-mod}%
The security model of our scheme  is defined by the following three games
executed between an adversary $\mathcal{A}$ and a challenger $\mathcal{C}$.
\medskip

\subsection{Unlinkability Game}%
\label{unlink-game}%
This game is used to define the unlinkability, \ie even if some ticket verifiers
collude with  potential users, they cannot profile the whole service information
of other users. We assume that $\mathcal{S}$ and $\mathcal{P}$ cannot be
compromised because  they can know a user's whole service information by
themselves. This game is formalized as follows: \medskip

\noindent{\sf Setup.} $\mathcal{C}$ runs {\sf Setup}$(1^{\ell})\rightarrow (MSK,PP)$ and sends $PP$ to $\mathcal{A}$.
\medskip

\noindent{\sf Phase 1.} $\mathcal{A}$ can make the following queries. 
\medskip

\noindent{\sf Registration Query.} $\mathcal{A}$ adaptively makes the following registration queries.
\begin{enumerate}
\item{\sf Ticket Issuer Registration Query.}  $\mathcal{C}$ runs $\mathcal{KG}(1^{\ell})\rightarrow(SK_{I},PK_{I})$ and {\sf Ticket-Issuer-Registration} $(\mathcal{I}(ID_{I},SK_{I},$ $PK_{I},PP)\leftrightarrow \mathcal{CA}(MSK,PK_{S},PP))\rightarrow(\sigma_{I},(ID_{I}, PK_{I}))$, and sends $(PK_{I},$ $\sigma_{I})$ to $\mathcal{A}$.
\medskip
\item{\sf Ticket Verifier Registration Query.} Let $Corrupt_{V}$ be the set consisting of the identities of verifiers corrupted by $\mathcal{A}$. $\mathcal{A}$  can adaptively submit a verifier' identity  $ID_{V}$: (1) if $ID_{V}\in Corrupt_{V}$,  $\mathcal{A}$ first runs $\mathcal{KG}(1^{\ell})\rightarrow(SK_{V},PK_{V})$, and then runs {\sf Ticket-Verifier-Registration}$(\mathcal{V}(ID_{V},SK_{V},PK_{V},PP)$ $\leftrightarrow \mathcal{CA}(MSK,PK_{V},$ $PP))\rightarrow(\sigma_{V},(ID_{V},PK_{V}))$ with $\mathcal{C}$; (2) if $ID_{V}\notin Corrupt_{V}$, $\mathcal{C}$ runs $\mathcal{KG}(1^{\ell})\rightarrow(SK_{V},PK_{V})$, and {\sf Ticket-Verifier-Registration}$(\mathcal{V}(ID_{V},SK_{V},PK_{V},PP)\leftrightarrow \mathcal{CA}(MSK,$ $PK_{V},PP))\rightarrow(\sigma_{V},(ID_{V},PK_{V}))$, and sends $(PK_{V},\sigma_{V})$
to $\mathcal{A}$. 
\medskip
\item{\sf User Registration Query.} $\mathcal{A}$ can adaptively submit a user's identity $ID_{U}$ and runs $\mathcal{KG}(1^{\ell})\rightarrow(SK_{U},PK_{U})$.  $\mathcal{A}$ and $\mathcal{C}$ run  {\sf User-Registration}$(\mathcal{U}(ID_{U},SK_{U},PK_{U},PP)\leftrightarrow \mathcal{CA}(MSK,PK_{U},PP))$ $\rightarrow(\sigma_{S},(ID_{U},PK_{U}))$.  $\mathcal{C}$ sends $\sigma_{U}$ to $\mathcal{A}$.
\medskip
\item{\sf Central Verifier Registration Query.}  $\mathcal{C}$ runs $\mathcal{KG}\rightarrow (SK_{CV},PK_{CV})$ and  {\sf Central-Verifier-Registration} $(\mathcal{CV}(ID_{CV},SK_{CV},PK_{CV},PP)\leftrightarrow \mathcal{CA}(MSK,PK_{CV},PP))\rightarrow(\sigma_{CV},(ID_{CV},PK_{CV}))$. $\mathcal{C}$ sends $(PK_{CV},$ $\sigma_{CV})$ to $\mathcal{A}$.  
\end{enumerate}
\medskip

\noindent{\sf Ticket Issuing Query.}  $\mathcal{A}$  adaptively submits  a set of {service} information $J_{U}$ to $\mathcal{C}$. $\mathcal{C}$ runs {\sf Ticket-Issuing}$(\mathcal{U}(SK_{U},$ $PK_{U},J_{U},\sigma_{U},PP)\leftrightarrow \mathcal{I}(SK_{I},PK_{I},\sigma_{I},PP))\rightarrow (T_{U},J_{U})$. Let $QI$ be the set which consists of the ticket information queried by $\mathcal{A}$ and initially empty. $\mathcal{C}$ adds $(T_{U},J_{U})$ into $QI$ and sends $T_{U}$ to $\mathcal{A}$. 
\medskip

\noindent{\sf Ticket Validation Query.} { $\mathcal{C}$ initializes a table $T_{V}$.  $\mathcal{A}$ can adaptively submit an authentication tag $Tag_{V}$ to $\mathcal{C}$. If $Tag_{V}\in T_{V}$, $\mathcal{C}$ aborts; otherwise, $\mathcal{C}$ adds $Tag_{V}$ into $T_{V}$ and works as follows.} $\mathcal{C}$ runs  {\sf Ticket-Validating}$(\mathcal{U}(SK_{U},PK_{U}, $ $Tag_{V},PP)\leftrightarrow \mathcal{V}(SK_{V},PK_{V},PK_{I},PP))\rightarrow(\bot,(1,Tag_{V})/$ $(0,Tag_{V}))$ and returns $(1,Tag_{V})$ to $\mathcal{A}$ if $Tag_{V}$ is valid and {$V\in J_{U}$}; otherwise, $(0,Tag_{V})$ is returned to indicate ${V}\notin J_{U}$. Let $QV$ be the set which consists of the ticket validation  queried by $\mathcal{A}$ and initially empty. $\mathcal{C}$ adds $(T_{U},J_{U})$ into $QV$.
\medskip

\noindent{\sf Ticket Trace Query.} $\mathcal{A}$ can adaptively submit a ticket $T_{U}$. $\mathcal{C}$  runs {\sf Ticket-Trace}$(\mathcal{U}(T_{U})\leftrightarrow \mathcal{CV}(SK_{CV}, $ $PK_{CV},T_{U},PP))\rightarrow(\bot,J_{U})$, and returns $J_{U}$ to $\mathcal{A}$ if $T_{U}\in QI$. Let $QT$ be the set which consists of the ticket trace information queried by $\mathcal{A}$ and initially empty. $\mathcal{C}$ adds $(T_{U},J_{U})$ into $QT$.
\medskip

\noindent{\sf Challenge.} { $\mathcal{A}$ submits two verifiers $V_{0}^{*}$ and $V_{1}^{*}$ with the limitation that $ID_{V_{0}^{*}},ID_{V_{1}^{*}}\notin Corrupt_{V}$. $\mathcal{C}$ flips an unbiased coin with $\{0,1\}$ and obtains a bit $b\in\{0,1\}$.  $\mathcal{C}$ sets $J_{U^{*}}=\{V_{b}^{*}\}$ and runs {\sf Ticket-Issuing}$(\mathcal{U}(SK_{U^{*}},PK_{U^{*}},J_{U^{*}},$ $\sigma_{U^{*}},PP)\leftrightarrow \mathcal{I}(SK_{I},PK_{I},\sigma_{I},PP))\rightarrow (T_{U^{*}},J_{U^{*}})$ where $T_{U^{*}}=(Tag_{V_{b}}^{*},Tag_{CV})$ and $Tag_{V_{b}}^{*}\notin T_{U}$ for all $(T_{U},J_{U})\in QI$, $(T_{U},J_{U})\in QV$  and $(T_{U},J_{U})\in QT$.  $\mathcal{C}$ sends $T_{U^{*}}$ to $\mathcal{A}$.}
\medskip

\noindent{\sf Phase 2.} It is the same as in {\sf Phase 1}.
\medskip

\noindent{\sf Output.} $\mathcal{A}$ outputs his guess $b'$ on $b$. $\mathcal{A}$ wins the game if $b'=b$.

\begin{definition}
An anonymous single-sign-on for $n$ designated services with
traceability scheme is $(\varrho_{1},\varrho_{2},\varrho_{3},$ $\epsilon(\ell))$ user secure if for all probabilistic polynomial-time  (PPT) adversary $\mathcal{A}$ making at moset $\varrho_{1}$ ticket issuing queries, $\varrho_{2}$ ticket validation queries and $\varrho_{3}$ ticket trace queries can win the above game with negligible advantage, namely
$Adv_{\mathcal{A}}=\left|\Pr\left[b'=b\right]-\frac{1}{2}\right|\leq \epsilon(\ell).$
\end{definition}

We say that a    scheme is selectively unlinkable if an initialization phase {\sf Initialization} is added before the the {\sf Setup} phase.
\medskip

\subsection{Unforgeability Game}%
\label{unforge-game}%
This game is used to define the unforgeability of tickets, namely even if users,
verifiers and the central verifier collude, they cannot forge a valid ticket.
This game is formalized as follows: \medskip

\noindent{\sf Setup.} $\mathcal{C}$ runs {\sf Setup}$(1^{\ell})\rightarrow 
(MSK,PP)$ and sends $PP$ to $\mathcal{A}$.
\medskip

\noindent{\sf Registration Query.} $\mathcal{A}$ can  make the following 
queries.
\begin{enumerate}

\item{\sf Ticket Seller Registration Query.}  $\mathcal{C}$ runs
$\mathcal{KG}(1^{\ell})\rightarrow(SK_{I},PK_{I})$ and {\sf
Ticket-Seller-Registration} $(\mathcal{I}(ID_{I},SK_{I},$
$PK_{I},PP)\leftrightarrow
\mathcal{CA}(MSK,PK_{I},PP))\rightarrow(\sigma_{I},(ID_{I}, $ $PK_{I}))$, and
sends $(PK_{I},\sigma_{I})$ to $\mathcal{A}$. %
\medskip

\item{\sf Ticket Verifier Registration Query.} $\mathcal{A}$ submits an identity
$ID_{V}$ and runs $\mathcal{KG}(1^{\ell})\rightarrow(SK_{V},PK_{V})$.
$\mathcal{A}$ and $\mathcal{C}$ run {\sf Ticket-Verifier-Registration}
$(\mathcal{V}(ID_{V},SK_{V},PK_{V},PP)\leftrightarrow
\mathcal{CA}(MSK,PK_{V},PP))\rightarrow(\sigma_{V},(ID_{V},$ $PK_{V}))$.
$\mathcal{C}$ returns $\sigma_{V}$ to $\mathcal{V}$;%
\medskip

\item{\sf User Registration Query.}  $\mathcal{A}$ submits an identity $ID_{U}$
and runs $\mathcal{KG}(1^{\ell})\rightarrow(SK_{U},PK_{U})$.   $\mathcal{A}$ and
$\mathcal{C}$  run {\sf User-Registration}$(\mathcal{U}(ID_{U},SK_{U},$
$PK_{U},PP)\leftrightarrow
\mathcal{CA}(MSK,PK_{U},PP))\rightarrow(\sigma_{S},(ID_{U},PK_{U}))$.
$\mathcal{C}$ returns $\sigma_{U}$. 
\medskip

\item{\sf Central Verifier Registration Query.} $\mathcal{A}$ submits a central
verifier's identity $ID_{CV}$ and  runs $\mathcal{KG}(1^{\ell})\rightarrow
(SK_{CV},PK_{CV})$. $\mathcal{A}$ and $\mathcal{C}$  run  {\sf
Central-Verifier-Registration}
$(\mathcal{CV}(ID_{CV},SK_{CV},PK_{CV},PP)\leftrightarrow \mathcal{CA}(MSK,$
$PK_{CV},PP))\rightarrow(\sigma_{CV},(ID_{CV},PK_{CV}))$. $\mathcal{C}$ sends
$\sigma_{CV}$ to $\mathcal{A}$
\end{enumerate}

\noindent{\sf Ticket Issuing Query.}  $\mathcal{A}$ adaptively submits  a set of
service information $J_{U}$. $\mathcal{C}$ runs {\sf Ticket-Issuing}
$\big(\mathcal{U}$ $(SK_{U},PK_{U},J_{U},\sigma_{U},PP)\leftrightarrow
\mathcal{I}(SK_{I},PK_{I},\sigma_{I},PP)\big)\rightarrow (T_{U}, J_{U})$ and
sends $T_{U}$ to $\mathcal{A}$. Let $QI$ be the set which consists of the ticket
information queried by $\mathcal{A}$ and initially empty. $\mathcal{C}$ adds
$(T_{U},J_{U})$ into $QI$. \medskip

\noindent{\sf Output.} $\mathcal{A}$ outputs a ticket
$T_{U^{*}}=\{Tag_{V^{*}}|V^{*} \in J_{U^{*}}\}\cup \{Tag_{CV}\}$ for a user
$\mathcal{U}^{*}$  with a set of service information $J_{U^{*}}$. $\mathcal{A}$
wins the game if {\sf Ticket-Validating}
$(\mathcal{U}(SK_{U^{*}},PK_{U^{*}},Tag_{V^{*}},PP) \leftrightarrow
\mathcal{V}((SK_{V^{*}},$ $PK_{V^{*}}),PK_{I},$
$PP))\rightarrow(\bot,(1,Tag_{V^{*}}))$ for all ${V^{*}\in J_{U^{*}}}$ and
$(T_{U^{*}},J_{U^{*}})\notin QI$.

\begin{definition}%
An anonymous single-sign-on for $n$ designated services with traceability is
$(\varrho,\epsilon(\ell))$ ticket-seller secure if for all probabilistic
polynomial-time (PPT) adversaries $\mathcal{A}$ who make $\varrho$  ticket
issuing queries can only win the above game with negligible advantage, namely
\begin{equation*}%
Adv_{\mathcal{A}}=\Pr\left[\begin{array}{cc}{\mbox{\sf 
Ticket-Validating}}(\mathcal{U}(SK_{U^{*}},PK_{U^{*}},Tag_{V^{*}},PP)\leftrightarrow
 \\
\mathcal{V}((SK_{V^{*}},PK_{V^{*}}),PK_{I},PP))
\rightarrow(\bot,(1,Tag_{V^{*}}))
\end{array}\right]\leq \epsilon(\ell)
\end{equation*}
for all ${V^{*}}\in J_{U^{*}}$.
\end{definition}

\subsection{Traceability Game}%
\label{trace-game}%

This game is used to formalize the traceability of tickets, namely even if a
group of users  collude, they cannot generate a ticket which the {\sf Ticket
Trace} algorithm would not catch as belonging to some member of the colluding
group. We suppose that the ticket issuer is honest. This game is formalized as
follows. 
\medskip\\ 

\noindent{\sf Setup.} $\mathcal{C}$ runs {\sf Setup}$(1^{\ell})\rightarrow
(MSK,PP)$ and sends $PP$ to $\mathcal{A}$. \medskip

\noindent{\sf Registration Query.} $\mathcal{A}$ can  make the following queries.
\begin{enumerate}
\item{\sf Ticket Issuer Registration Query.}  $\mathcal{C}$ runs
$\mathcal{KG}(1^{\ell})\rightarrow(SK_{S},PK_{S})$ and {\sf
Ticket-Issuer-Registration} $(\mathcal{I}(ID_{I},SK_{I},$
$PK_{I},PP)\leftrightarrow
\mathcal{CA}(MSK,PK_{S},PP))\rightarrow(\sigma_{I},(ID_{I}, $ $PK_{I}))$, and
sends $(PK_{I},\sigma_{I})$ to $\mathcal{A}$. \medskip \item{\sf Ticket Verifier
Registration Query.} $\mathcal{A}$ selects an verifier $V$ and runs
$\mathcal{KG}(1^{\ell})\rightarrow(SK_{V},PK_{V})$.  $\mathcal{A}$ and
$\mathcal{C}$ runs {\sf Ticket-Verifier-Registration}
$(\mathcal{V}(ID_{V},SK_{V},PK_{V},PP)\leftrightarrow
\mathcal{CA}(MSK,PK_{V},PP))\rightarrow(\sigma_{V},(ID_{V},$ $PK_{V}))$.
$\mathcal{C}$ sends $\sigma_{V}$ to $\mathcal{A}$. \medskip
 
\item{\sf User Registration Query.}  $\mathcal{A}$ selects an
identity $ID_{U}$ and runs
$\mathcal{KG}(1^{\ell})\rightarrow(SK_{U},PK_{U})$.  $\mathcal{A}$ and $\mathcal{C}$ runs {\sf
	User-Registration}$(\mathcal{U}(ID_{U},SK_{U},$ $PK_{U},PP)\leftrightarrow
\mathcal{CA}(MSK,PK_{U},PP))\rightarrow(\sigma_{S},(ID_{U},PK_{U}))$. $\mathcal{C}$ sends $\sigma_{U}$ to
$\mathcal{A}$.  Let $QK_{U}$ be the set which consists
 of the users' identities selected by $\mathcal{A}$ to make registration query
 and is initially empty. 
\medskip

\item{\sf Central Verifier Registration Query.} $\mathcal{C}$ selects a central
verifier's identity $ID_{CV}$, runs
$\mathcal{KG}(1^{\ell})\rightarrow(SK_{CV},PK_{CV})$ and  {\sf
Central-Verifier-Registration}
$(\mathcal{CV}(ID_{CV},SK_{CV},PK_{CV},PP)\leftrightarrow \mathcal{CA}(MSK,$
$PK_{CV},PP))\rightarrow(\sigma_{CV},(ID_{CV},PK_{CV}))$.  $\mathcal{C}$ sends
$(PK_{CV},\sigma_{CV})$ to $\mathcal{A}$.
\end{enumerate}

\noindent{\sf Ticket Issuing Query.}  $\mathcal{A}$ adaptively submits  a set of
service information $J_{U}$. $\mathcal{C}$ runs {\sf Ticket-Issuing}
$\big(\mathcal{U}$ $(SK_{U},$ $PK_{U},J_{U},\sigma_{U},PP)\leftrightarrow
\mathcal{I}(SK_{I},PK_{I},\sigma_{I},PP)\big)\rightarrow (T_{U}, J_{U})$ and
sends $T_{U}$ to $\mathcal{A}$. Let $QI$ be the set which consists of the ticket
information queried by $\mathcal{A}$ and initially empty. $\mathcal{C}$ adds
$(T_{U},J_{U})$ into $QI$. \medskip

\noindent{\sf Output.} $\mathcal{A}$ outputs a ticket
$T_{U^{*}}=\{Tag_{V^{*}}|V^{*}\in J_{U^{*}}\}\cup \{Tag_{CV}^{*}\}$ for a user
$\mathcal{U}^{*}$  with a set of service information $J_{U^{*}}$. $\mathcal{A}$
wins the game if {\sf Ticket-Trace}
$((SK_{CV},PK_{CV},Tag_{CV}^{*},T_{U}^{*},PP))\rightarrow
(\tilde{\mathcal{U}},J_{\tilde{U}}))$ with $\tilde{\mathcal{U}}\notin QK_{U} $
or $\mathcal{U}^{*}\neq\tilde{\mathcal{U}}\in QK_{U}$.

\begin{definition}
An anonymous single-sign-on for $n$ designated services with
traceability scheme is
$(\varrho,\epsilon(\ell))$ traceable if for all probabilistic
polynomial-time (PPT) adversaries $\mathcal{A}$ who make $\varrho$  ticket
issuing queries can only win the above game with negligible advantage, namely
\begin{equation*}%
Adv_{\mathcal{A}}=\Pr\left[\begin{array}{c|c}\tilde{\mathcal{U}}\notin QK_{U} 
~\mbox{or}&\mbox{{\sf Ticket-Trace}}(SK_{CV},PK_{CV},\\
~ \mathcal{U}^{*}\neq\tilde{\mathcal{U}}\in QK_{U} 
&Tag_{CV}^{*},T_{U}^{*},PP)\rightarrow (\mathcal{U}',J_{U'})
\end{array}\right]\leq \epsilon(\ell).
\end{equation*}
\end{definition}

\section{The Detail of $\prod_{U}^{1}$ }\label{d1}

An instantiation of the proof $\prod_{U}^{1}$ is as follows. 
$\mathcal{U}$ select $v_{1},v_{2},z_{u},r'_{u},x'_{u},e_{u}',v_{2}',v'_{3},v',z'_{1},z'_{2},$ $\cdots,z'_{n}\stackrel{R}{\leftarrow}\mathbb{Z}_{p}$ and computes $v_{3}=\frac{1}{v_{1}}$,  $v=r_{u}-v_{2}v_{3}$,  $\bar{\sigma}_{U}=\sigma_{U}^{v_{1}}$,  $\tilde{\sigma}_{U}=\bar{\sigma}_{U}^{-e_{u}}B_{U}^{v_{1}}(=\bar{\sigma}_{U}^{x_{a}})$,  $\bar{B}_{U}=B_{U}^{v_{1}}h^{-v_{2}}$,  $W_{1}=\bar{\sigma}_{U}^{-e'_{u}}h^{v'_{2}}$, $W_{2}=\bar{B}_{U}^{-v'_{3}}\xi^{x'_{u}}h^{v'}$,   $\big(z_{v}=H_{1}(z_{u}||ID_{V}),$  $P_{V}=Y_{U}{Y}_{P}^{z_{v}}, P'_{v}=\xi^{x_{u}}{Y}_{P}^{z'_{v}}, Q_{V}=\xi^{z_{v}}, Q'_{V}=\xi^{z'_{v}}\big)_{ID_{V}\in J_{U}}$. $\mathcal{U}$ computes $c=H_{1}(\bar{\sigma}_{U}||\tilde{\sigma}_{U}||\bar{B}_{U}||W_{1}||W_{2}||P_{1}||P_{1}'||Q_{1}||Q'_{1}||$ $P_{2}||P_{2}'||Q_{2}||Q'_{2}||\cdots||P_{n}||P_{n}'||$ $Q_{n}||Q'_{n})$,  
$\hat{e}_{u}=e'_{u}-ce_{u}$, $\hat{v}_{2}=v'_{2}-cv_{2}$, $\hat{v}_{3}=v'_{3}-cv_{3}$, $\hat{v}=v'-cv$, $\hat{x}_{u}=x'_{u}-cx_{u}$ and $(\hat{z}_{v}=z'_{v}-cz_{v})_{ID_{V}\in J_{U}}$.
 $\mathcal{U}$ sends $(\bar{\sigma}_{U},\tilde{\sigma}_{U},\bar{B}_{U},W_{1},W_{2},(P_{V},P'_{V},Q_{V},Q'_{V})_{ID_{V}\in J_{U}})$ and $(c,\hat{e}_{u},$\\ $\hat{v}_{2},\hat{v}_{3},\hat{v},\hat{x}_{u},\hat{z}_{1},\hat{z}_{2},\cdots, \hat{z}_{n})$ to $\mathcal{S}$.
 \medskip
 
After receiving 
$(\bar{\sigma}_{U},\tilde{\sigma}_{U},\bar{B}_{U},W_{1},W_{2},(P_{V},P'_{V},Q_{V},Q'_{V})_{ID_{V}\in
 J_{U}})$ and 
$(c,\hat{e}_{u},\hat{v}_{2},\hat{v}_{3},\hat{v},\hat{x}_{u},\hat{z}_{1},$\\ 
$\hat{z}_{2},\cdots,\hat{z}_{n})$, $\mathcal{S}$ checks
\begin{equation*}
\begin{split}
c  \stackrel{?}{=}   
H_{1}(\bar{\sigma}_{U}||\tilde{\sigma}_{U}||\bar{B}_{U}||W_{1}||W_{2}||P_{1}||P_{1}'||Q_{1}||Q'_{1}||P_{2}||P_{2}'||Q_{2}||Q'_{2}||\cdots||P_{n}||P_{n}'||Q_{n}||Q'_{n}),
 \\ 
W_{1}\stackrel{?}{=}  
\bar{\sigma}_{U}^{-\hat{e}_{u}}h^{\hat{v}_{2}}(\frac{\tilde{\sigma}_{U}}{\bar{B}_{U}})^{c},
 ~ W_{2}\stackrel{?}{=}  
\bar{B}_{U}^{-\hat{v}_{3}}\xi^{\hat{x}_{u}}h^{\hat{v}}g^{-c},  ~ 
(P'_{V}\stackrel{?}{=}   \xi^{\hat{x}_{u}}Y_{P}^{\hat{z}_{v}}P_{V}^{c}, ~
 Q'_{V}\stackrel{?}{=}   \xi^{\hat{z}_{v}} Q_{V}^{c})_{ID_{V}\in J_{U}}.
 \end{split}
 \end{equation*}

\section{The Detail of $\prod_{U}^{2}$ }\label{d2}
An instantiation of the proof $\prod_{U}^{2}$ is as follows. $\mathcal{U}$ selects $x'_{u},z'_{v}\stackrel{R}{\leftarrow}\mathbb{Z}_{p}$, and computes $P_{V}'=$\\ $\xi^{x'_{u}}Y_{P}^{z'_{v}}$, $Q'_{V}=\xi^{z'_{v}}$, $c_{v}=H_{1}(P_{V}||P'_{V}||Q_{V}||Q'_{V})$, $\hat{x}_{u}=x'_{u}-c_{v}x_{u}$ and $\hat{z}_{v}=z'_{v}-c_{v}z_{v}$. $\mathcal{U}$\\ sends $(P_{V},P'_{V},Q_{V},Q'_{V})$ and $(c_{v},\hat{x}_{v},\hat{z}_{v})$ to $\mathcal{V}$.
\medskip

After receiving $(P_{V},P'_{V},Q_{V},Q'_{V})$ and $(c_{v},\hat{x}_{v},\hat{z}_{v})$, $\mathcal{V}$ verifiers 
\begin{equation*}
c_{v}\stackrel{?}{=}H_{1}(P_{V}||P'_{V}||Q_{V}||Q'_{V}), ~P'_{V}\stackrel{?}{=}\xi^{\hat{x}_{u}}Y_{P}^{\hat{z}_{v}}P_{V}^{c_{v}}~\mbox{and}~Q'_{V}\stackrel{?}{=}\xi^{\hat{z}_{v}}Q_{V}^{c_{v}}.
\end{equation*}

\section{Correctness}\label{correct}
Our scheme is correct as the following equations hold.

\begin{equation*}
\begin{split}
e(\sigma_{I},Y_{A}\mathfrak{g}^{e_{i}})=e((gh^{r_{i}}Y_{I})^{\frac{1}{x_{a}+e_{i}}},\mathfrak{g}^{x_{a}+e_{i}})=e(gh^{r_{i}}Y_{I},\mathfrak{g}),\\
e(\sigma_{V},Y_{A}\mathfrak{g}^{e_{v}})=e((gh^{r_{v}}Y_{V})^{\frac{1}{x_{a}+e_{v}}},\mathfrak{g}^{x_{a}+e_{v}})=e(gh^{r_{v}}Y_{V},\mathfrak{g}),\\
\end{split}
\end{equation*}
\begin{equation*}
\begin{split}
e(\sigma_{U},Y_{A}\mathfrak{g}^{e_{u}})=e((gh^{r_{u}}Y_{U})^{\frac{1}{x_{a}+e_{u}}},\mathfrak{g}^{x_{a}+e_{u}})=e(gh^{r_{u}}Y_{U},\mathfrak{g}),\\
e(\sigma_{CV},Y_{A}\mathfrak{g}^{e_{cv}})=e((gh^{r_{cv}}Y_{CV})^{\frac{1}{x_{a}+e_{cv}}},\mathfrak{g}^{x_{a}+e_{cv}})=e(gh^{r_{cv}}Y_{CV},\mathfrak{g}),\\
\end{split}
\end{equation*}

\begin{equation*}
\begin{split}
\tilde{\sigma}_{U}&=\bar{\sigma_{U}}^{-e_{u}}B_{U}^{v_{1}}=\sigma_{U}^{-e_{u}v_{1}}B_{U}^{v_{1}}=B_{U}^{\frac{-e_{u}v_{1}}{x_{a}+e_{u}}}B_{U}^{v_{1}}=B_{U}^{\frac{-v_{1}(e_{u}+x_{a})+v_{1}x_{a}}{x_{a}+e_{u}}}B_{U}^{v_{1}}=\\
&B_{U}^{-v_{1}}B_{U}^{\frac{v_{1}x_{a}}{x_{a}+e_{u}}}B_{U}^{v_{1}}=(B_{U}^{\frac{1}{x_{a}+e_{u}}})^{v_{1}x_{a}}=(\sigma_{U}^{v_{1}})^{x_{a}}=\bar{\sigma}_{U}^{x_{a}},
\end{split}
\end{equation*}

\begin{equation*}
\begin{split}
\frac{\tilde{\sigma}_{U}}{\bar{B}_{U}}=\frac{\bar{\sigma}_{U}^{-e_{u}}B_{U}^{v_{1}}}{B_{U}^{v_{1}}h^{-v_{2}}}=\bar{\sigma}_{U}^{-e_{u}}h^{v_{2}},
\end{split}
\end{equation*}

\begin{equation*}
\begin{split}
\bar{B}_{U}^{-v_{3}}\xi^{x_{u}}h^{v}=(B_{U}^{v_{1}}h^{-v_{2}})^{-v_{3}}\xi^{x_{u}}h^{v}=((gh^{r_{u}}Y_{U})^{v_{1}}h^{-v_{2}})^{-v_{3}}\xi^{x_{u}}h^{v}=\\
(gh^{r_{u}}Y_{U})^{-1}h^{v_{2}v_{3}}\xi^{x_{u}}h^{v}=g^{-1}h^{-r_{u}}Y_{U}^{-1}Y_{U}h^{v_{2}v_{3}+v}=g^{-1}h^{v_{2}v_{3}-r_{u}+v}=g^{-1},
\end{split}
\end{equation*}
\begin{equation*}
\begin{split}
e(\sigma_{V},\tilde{Y}_{I}\mathfrak{g}^{e_{v}})=e((gh^{w_{v}}\tilde{h}^{s_{v}})^{\frac{1}{x_{i}+e_{v}}},\mathfrak{g}^{x_{i}+e_{v}})
=e(gh^{w_{v}}\tilde{h}^{s_{v}},\mathfrak{g}),
\end{split}
\end{equation*}
\begin{equation*}
\begin{split}
e(\sigma_{CV},\tilde{Y}_{I}\mathfrak{g}^{e_{cv}})=e((gh^{w_{cv}}\tilde{h}^{s_{cv}})^{\frac{1}{x_{i}+e_{cv}}},\mathfrak{g}^{x_{i}+e_{cv}})
=e(gh^{w_{cv}}\tilde{h}^{s_{cv}},\mathfrak{g}),
\end{split}
\end{equation*}
\begin{equation*}
\begin{split}
E_{V}^{x_{v}}=\xi^{x_{v}d_{v}}=Y_{V}^{d_{v}}=F_{V},
\end{split}
\end{equation*}
\begin{equation*}
\begin{split}
\frac{P_{V}}{Q_{V}^{x_{cv}}}=\frac{Y_{U}Y_{CV}^{z_{v}}}{\xi^{x_{cv}z_{v}}}=\frac{Y_{U}Y_{CV}^{z_{v}}}{Y_{CV}^{z_{v}}}=Y_{U},
\end{split}
\end{equation*}
and 
\begin{equation*}
\begin{split}
\frac{K_{V}}{E_{V}^{x_{cv}}}=\frac{Y_{V}Y_{CV}^{d_{v}}}{\xi^{x_{cv}d_{v}}}=\frac{Y_{V}Y_{CV}^{d_{v}}}{Y_{CV}^{d_{v}}}=Y_{V}.
\end{split}
\end{equation*}

\section{Proof of Theorem \ref{theorem:unlink}}%
\label{app:unlink}
\begin{proof} If there exists an adversary $\mathcal{A}$ can $(\varrho_{1},\varrho_{2},\varrho_{3},\epsilon'(\ell))$ break the selective unlinkability of our scheme, we can construct an algorithm $\mathcal{B}$ which can use $\mathcal{A}$ as a subroutine to break the decisional  Diffie-Hellman (DDH) assumption as follows. Given $(\xi,\xi^{\alpha},\xi^{\beta})$, $\mathcal{C}$ flips an unbiased coin with $\{0,1\}$, and obtains a bit $b\in\{0,1\}$. If $b=0$, 
$\mathcal{C}$ sends $T=\xi^{\alpha\beta}$ to $\mathcal{B}$; If $b=1$, 
$\mathcal{C}$ sends $T=M$ to $\mathcal{B}$, where 
$M\stackrel{R}{\leftarrow}\mathbb{G}_{2}$. $\mathcal{B}$ will output his guess 
$b'$ on $b$.
\medskip

\noindent{\sf Initialisation.} $\mathcal{A}$ submits two verifiers  ${V_{0}^{*}}$ and ${V_{1}^{*}}$.   $\mathcal{B}$ flip an unbiased coin with $\{0,1\}$ and obtains a bit $\mu\in\{0,1\}$.  $\mathcal{B}$ sets $Y_{V_{\mu}^{*}}=\xi^{\alpha}$ and  $Y_{V_{1-\mu}^{*}}=\xi^{\gamma}$ where $\gamma\stackrel{R}{\leftarrow}\mathbb{Z}_{p}$. 
\medskip

\noindent{\sf Setup.} $\mathcal{B}$ selects  $x_{a}\stackrel{R}{\leftarrow}\mathbb{Z}_{p}$, $g,h,\xi,\tilde{h},\stackrel{R}{\leftarrow}\mathbb{G}_{1}$ and   $\mathfrak{g}\stackrel{R}{\leftarrow}\mathbb{G}_{2}$. $\mathcal{B}$ computes $Y_{A}=\mathfrak{g}^{x_{a}}$, and selects $H_{1}:\{0,1\}^{*}\rightarrow\mathbb{Z}_{p}$ and  $H_{2}:\left\{0,1\right\}^{*}\rightarrow \mathbb{Z}_{p}$. $\mathcal{B}$ sends the public parameters  $PP=(e,p,\mathbb{G}_{1},\mathbb{G}_{2},\mathbb{G}_{\tau},g,h,\xi,$ $\tilde{h},\mathfrak{g},Y_{A},H_{1},H_{2})$ to $\mathcal{A}$.
\medskip

\noindent{\sf Phase 1.} $\mathcal{A}$ can make the following  queries.
\medskip

\noindent{Registration Query.} $\mathcal{A}$ can make the following registration queries.
\begin{enumerate}
\item{\sf Ticket Issuer Registration Query.} $\mathcal{B}$ selects $x_{i},e_{i},r_{i}\stackrel{R}{\leftarrow}\mathbb{Z}_{p}$, and computs $Y_{I}=\xi^{x_{i}}$,  $\tilde{Y}_{I}=\mathfrak{g}^{x_{i}}$ and $\sigma_{I}=(gh^{r_{i}}Y_{I})^{\frac{1}{x_{a}+e_{i}}}$. $\mathcal{B}$ sends $(r_{i},e_{i},\sigma_{I},Y_{I},\tilde{Y}_{I})$ to $\mathcal{A}$.\medskip

\item{\sf Ticket Verifier Registration Query.} Let $Corrupt_{V} $  be the set consisting of the identities of verifiers corrupted by $\mathcal{A}$. $\mathcal{A}$ selects an identity $ID_{V}\notin\{ID_{V_{0}^{*}},ID_{V_{1}^{*}}\}$: (1) if $ID_{V}\in Corrupt_{V}$, $\mathcal{A}$ sends $(ID_{V}, Y_{V})$ to $\mathcal{B}$ where $Y_{V}$ is the public key of $ID_{V}$.  $\mathcal{B}$ selects $\lambda_{v},r_{v}\stackrel{R}{\leftarrow}\mathbb{Z}_{p}$, and computes $\sigma_{V}=(gh^{r_{v}}Y_{V})^{\frac{1}{x_{a}+e_{v}}}$. $\mathcal{B}$ sends $(\lambda_{v},r_{v},\sigma_{V})$ to $\mathcal{A}$; If  $ID_{V}\notin Corrupt_{V}$, $\mathcal{B}$ selects $x_{v},\lambda_{v},r_{v}\stackrel{R}{\leftarrow}\mathbb{Z}_{p}$ and computes $Y_{V}=\xi^{x_{v}}$ and $\sigma_{V}=(gh^{r_{v}}Y_{V})^{\frac{1}{x_{a}+e_{v}}}$. $\mathcal{B}$  sends $(Y_{V},\lambda_{v},r_{v},\sigma_{V})$  to $\mathcal{A}$. 
\medskip

\item{\sf User Registration Query.}  Let  $RQ_{U}$ be the set consisting of the registration information of users. $\mathcal{A}$ selects an identity $ID_{U}$ and sends $(ID_{U},Y_{U})$ to $\mathcal{B}$.  $\mathcal{B}$ selects $\lambda_{u},r_{u}\stackrel{R}{\leftarrow}\mathbb{Z}_{p}$, and computes  $\sigma_{U}=(g_{0}h^{r_{u}}Y_{U})^{\frac{1}{x_{a}+\lambda_{u}}}$.  $\mathcal{B}$ sends $(r_{u},\lambda_{u},\sigma_{U})$ to $\mathcal{A}$. $\mathcal{B}$ adds $(ID_{U},Y_{U},\lambda_{u},r_{u},\sigma_{U})$ to $QK_{U}$.  $\mathcal{A}$ can adaptively make this registration queries multiple times.
\medskip

\item{\sf Central Verifier Registration Query.}  $\mathcal{CV}$ selects $x_{cv}, \lambda_{cv},r_{cv}\stackrel{R}{\leftarrow}\mathbb{Z}_{p}$ and computes $Y_{CV}=\xi^{x_{cv}}$ and $\sigma_{CV}=(gh^{r_{cv}}Y_{CV})^{\frac{1}{x_{a}+\lambda_{cv}}}$. $\mathcal{B}$ sends $(Y_{CV},r_{cv},\lambda_{cv},\sigma_{CV})$ to $\mathcal{A}$.
\end{enumerate}
\medskip

\noindent{\sf Ticket Issuing Query.} $\mathcal{A}$ submits an identity $ID_{U}\in QR_{U}$,  a set of {service} information $J_{U}$, a set of pseudonym $Ps_{U}=\{(P_{V},Q_{V})_{{V}\in J_{U}})\}$, and a proof  $\mbox{PoK}\{(x_{u},r_{u},e_{u},\sigma_{U},v_{1},$ $v_{2},v_{3},(z_{v})_{{V}\in J_{U}}): \frac{\tilde{\sigma}_{U}}{\bar{B}_{U}}=\bar{\sigma}_{U}^{-e_{u}}h^{v_{2}} \wedge g_{0}^{-1}=\bar{B}_{U}^{v_{3}}g^{x_{u}}h^{r_{u}-v_{2}v_{3}}\wedge(P_{V}=\xi^{x_{u}}{Y}_{P}^{z_{v}}\wedge Q_{V}=\xi^{z_{v}})_{{V}\in J_{U}}\}$. If the proof is incorrect, $\mathcal{B}$ aborts. Otherwise, $\mathcal{B}$ works as follows.

For ${V}\in J_{U}$, $\mathcal{B}$ selects $t_{u},d_{v},w_{v},e_{v}\stackrel{R}{\leftarrow}\mathbb{Z}_{p}$   and   computes
\begin{equation*} 
\begin{split} 
C_{U}=\xi^{t_{u}},~D_{V}=H_{2}(C_{U}||ID_{V}),~E_{V}=\xi^{d_{v}},~F_{V}= 
Y_{V}^{d_{v}},~K_{V}=Y_{V}{Y}_{P}^{d_{v}},\\
s_{v}=H_{1}(P_{V}||Q_{V}||E_{V}||F_{V}||K_{V}||Text)~\mbox{and}~\sigma_{V}= 
(gh^{w_{v}}\tilde{h}^{s_{v}})^{\frac{1}{x_{i}+e_{v}}}. 
\end{split}
\end{equation*}

$\mathcal{B}$ select $w_{cv},e_{cv}\stackrel{R}{\leftarrow}\mathbb{Z}_{p}$, and computes  $s_{cv}=H_{1}(s_{1}||s_{2}||\cdots||s_{|J_{U}|})$ and $Z_{CV}=(gh^{w_{cv}}\tilde{h}^{s_{cv}})^{\frac{1}{x+e_{cv}}}$. The ticket is
 $T_{U}=\big\{(D_{V},Tag_{V})|{V}\in J_{U}\big\} \cup \big\{s_{cv},w_{cv},e_{cv},Z_{CV}\big\}$. $\mathcal{B}$ returns $(C_{U},T_{U},Text)$ to $\mathcal{A}$. Let $QI$ be the set consisting of the tickets queried by $\mathcal{A}$ and is initially empty.  $\mathcal{B}$ adds { $\big\{Ps_{U},J_{U},T_{U},t_{u}\big\}\cup \big\{d_{v}|{V}\in J_{U}\big\}$ into $QI$. }
\medskip

\noindent{\sf Ticket Validation Query.} $\mathcal{B}$ initializes a table 
$T_{V}$.  $\mathcal{A}$ submits 
$Tag_{V}=(P_{V},Q_{V},E_{V},F_{V},K_{V},s_{v},w_{v},e_{v},$\\$\sigma_{V},Text)$ 
and a proof $\prod_{U}^{2}: \mbox{PoK}\{(x_{u},$ $z_{v}): 
P_{V}=\xi^{x_{u}}{Y}_{P}^{z_{v}}~\wedge~Q_{V}=\xi^{z_{v}}\}$. If $Tag_{V}\in 
T_{V}$, $\mathcal{B}$ aborts; otherwise, $\mathcal{B}$ adds $Tag_{V}$ into 
$T_{V}$ and works as follows.  $\mathcal{B}$ checks whether 
$(P_{V},Q_{V},E_{V},F_{V},K_{V},s_{v},w_{v},$ $\sigma_{V},Text)\in QI$. If not, 
$\mathcal{B}$ aborts; otherwise, $\mathcal{B}$ computes 
$Y_{V}=F_{j}^{\frac{1}{d_{v}}}$ and checks 
$D_{V}\stackrel{?}{=}H_{2}(C_{U}||ID_{V})$, 
$s_{v}\stackrel{?}{=}H_{1}(||P_{V}||Q_{V}||E_{V}$ $||F_{V}||K_{V}||Text)$ and  
$e(\sigma_{V},Y_{I}\mathfrak{g}^{e_{v}})\stackrel{?}{=} 
e(gh^{w_{v}}\tilde{h}^{s_{v}},\mathfrak{g})$. If the above equations hold. 
$\mathcal{B}$ returns $ID_{V}$ to $\mathcal{A}$; otherwise, $\perp$ is returned 
to indicate failure. Let $QV$ be the set consisting of ticket validation 
queries made by $\mathcal{A}$ and initially empty. $\mathcal{B}$ adds 
$(P_{V},Q_{V},E_{V},F_{V},K_{V},$ $s_{v},w_{v},e_{v},\sigma_{V})$  into $QV$. 
$\mathcal{A}$ can adaptively make this query up to $\rho_{2}$ times. 
\medskip

\noindent{\sf Ticket Trace Query.}   $\mathcal{A}$ submits a ticket $T_{U}$. 
$\mathcal{B}$ works as follows. (1) Let $\Omega_{U}=\{\}$. For each $Tag_V$ in 
$T_U$:  a) Compute: $Y_{U}=\frac{P_{V}}{Q_{V}^{x_{cv}}}$ and 
$Y_{V}=\frac{K_{V}}{E_{V}^{x_{cv}}}$. b) Look up $ID_V$ for $Y_V$'s. Check: 
(c1) $s_{v}\stackrel{?}{=}H_{1}(P_{V}||Q_{V}||E_{V}||K_{V}||Text)$; (c2) 
$e(Z_{V},Y_{I}\mathfrak{g}^{w_{v}})\stackrel{?}{=} 
e(gh^{w_{v}}\tilde{h}^{s_{v}},\mathfrak{g})$;
(d) If (c1) and (c2) hold, set $\Omega_U=\Omega_U\cup\{V\}$; otherwise abort. 
(e) verify $Y_U$ remains the same for all tags.   (2) 
$s_{cv}\stackrel{?}{=}H_{1}(s_{1}||s_{2}||\cdots||s_{|J_{U}|})$;  (3) 
$e(Z_{CV},\tilde{Y}_{I}\mathfrak{g}^{w_{cv}})\stackrel{?}{=}e(gh^{w_{cv}}\tilde{h}^{s_{cv}},\mathfrak{g})$.
 
If (1), (2) and (3) hold, $\mathcal{CV}$ can determine that the service 
information of $\mathcal{U}$ with public key $Y_{U}$ is:  $J_{U}=\Omega_U$; 
otherwise, the trace has failed. Let $QT$ be a set consisting of the ticket 
trace queries made by $\mathcal{A}$. $\mathcal{B}$ adds $T_{U}$ into $QT$.
\medskip

\noindent{\sf Challenge.} $\mathcal{B}$ selects 
$z_{\mu}^{*},t_{\mu}^{*},w_{\mu}^{*},e_{\mu}^{*}, 
w^{*},e^{*}\stackrel{R}{\leftarrow}\mathbb{Z}_{p}$, and computes 
$P_{\mu}^{*}=Y_{U}Y_{CV}^{z^{*}_{\mu}}$, $Q_{\mu}^{*}=\xi^{z^{*}_{\mu}}$, 
$C_{\mu}^{*}=\xi^{t_{\mu}^{*}}$, 
$D_{\mu}^{*}=H_{2}(C_{U}^{*}||Y_{V^{*}_{\mu}})$, $E_{\mu}^{*}=\xi^{\beta}$, 
$F_{\mu}^{*}=T$, $K_{\mu}^{*}=(E_{\mu}^{*})^{x_{cv}}Y_{V_{\mu}^{*}}$, 
$s_{\mu}^{*}=H_{1}(P_{U}^{*}||Q_{U}^{*}||E_{\mu}^{*}||F_{\mu}^{*}||K_{\mu}^{*}||$
 $Text)$, 
$\sigma_{\mu}^{*}= 
(gh^{w_{\mu}^{*}}\tilde{h}^{s_{\mu}^{*}})^{\frac{1}{x_{i}+e_{\mu}^{*}}}$,
$s^{*}=H_{1}(s_{\mu}^{*})$ and 
$\sigma^{*}=(gh^{w^{*}}\tilde{h}^{s^{*}})^{\frac{1}{x_{i}+e^{*}}}$. 
$\mathcal{B}$ sends 
$\{P_{\mu}^{*},Q_{\mu}^{*},E_{\mu}^{*},F_{\mu}^{*},K_{\mu}^{*},s_{\mu}^{*}, 
w_{\mu}^{*},$ $e_{\mu}^{*}, \sigma_{\mu}^{*}\}\cup 
\{s^{*},w^{*},e^{*},\sigma^{*}\}$ to 
$\mathcal{A}$. 
\medskip

\noindent{\sf Phase 2.} It is the same as in {\sf Phase 1} with the limitations: (1) $(E_{\mu}^{*},F_{\mu}^{*})\notin QV$; (2)  $(E_{\mu}^{*},F_{\mu}^{*})\notin QT$; (3) $\mathcal{A}$ can adaptively make the ticket issuing query, ticket validation query and ticket trace query at most $\varrho_{1}$, $\varrho_{2}$ and $\varrho_{3}$, respectively.
\medskip

\noindent{\sf Output.} $\mathcal{A}$ outputs his guess $\mu'$ on $\mu$. If $\mu'=\mu$, $\mathcal{B}$ outputs $b'=0$; otherwise, $\mathcal{B}$ outputs $b=1$.
\medskip

Now, we compute the probability with which $\mathcal{B}$ can break the DDH 
assumption. If $b=0$, $T=\xi^{\alpha\beta}$ and  
$(D_{\mu}^{*},P_{\mu}^{*},Q_{\mu}^{*},E_{\mu}^{*}, 
F_{\mu}^{*},K_{\mu}^{*},w_{\mu}^{*},e_{\mu}^{*}, \sigma_{\mu}^{*})$ is a valid 
authentication tag. Hence,  $\mathcal{A}$ can outputs $\mu'=\mu$ with   
$\Pr\left[\mu'=\mu|b=0\right]\geq\frac{1}{2}+\epsilon'(\ell)$. When $\mu'=\mu$, 
$\mathcal{B}$ outputs $b'=0$. We have, $\Pr\left[b'=b|b=0 \right]\geq 
\frac{1}{2}+\epsilon(\ell)$. If $b=1$, $T=M$ and 
$(D_{\mu}^{*},P_{\mu}^{*},Q_{\mu}^{*},E_{\mu}^{*},F_{\mu}^{*}, 
K_{\mu}^{*},w_{\mu}^{*},e_{\mu}^{*}, \sigma_{\mu}^{*})$ are random elements in 
$\mathbb{G}_{1}$. Hence, $\mathcal{A}$ can output $\mu'\neq\mu$ with 
$\Pr\left[\mu'\neq\mu|b=1\right]=\frac{1}{2}$. When $\mu'\neq\mu$, 
$\mathcal{B}$ outputs $b=1$. We have, $\Pr\left[b'=b|b=1\right]= \frac{1}{2}$. 
\medskip

Therefore, the advantage with which $\mathcal{B}$ can break the DDH assumption is 
\begin{equation*}
Adv_{\mathcal{B}}^{DDH}=\left|\frac{1}{2}\times 
\Pr\left[b'=b|b=0\right]-\frac{1}{2}\times \Pr\left[b'=b|b=1\right] \right|\geq 
\frac{1}{2}(\frac{1}{2}+\epsilon'(\ell))-\frac{1}{2}\times\frac{1}{2}= 
\frac{\epsilon'(\ell)}{2}.
\end{equation*}
\end{proof}

\section{Proof of Theorem \ref{theorem:unforge}}%
\label{app:unforge}
\begin{proof}%
If there exists an adversary $\mathcal{A}$ which can break the unforgeability of
our  scheme with the advantage $\epsilon'(\ell)$, we can constructs an algorithm
$\mathcal{B}$ which can use $\mathcal{A}$ as a subroutine to break the
JOC-$q$-SDH assumption as follows. Given a $(q+3)$-tuple
$(g,g^{x},\cdots,g^{x^{q}},\mathfrak{g},\mathfrak{g}^{x})$, $\mathcal{B}$ will
output $(c,g^{\frac{1}{x+c}})$ where $c\in\mathbb{Z}_{p}-\{-x\}$. \medskip

\noindent{\sf Setup.}  $\mathcal{B}$ selects
$e_{1},e_{2},\cdots,e_{q-1}\stackrel{R}{\leftarrow}\mathbb{Z}_{p}$, and sets
$f(x)=\prod_{i=1}^{q-1}(x+e_{i})=\sum_{i=0}^{q-1}\alpha_{i}x^{i}$,
$f_{i}(x)=\frac{f(x)}{x+e_{i}}=\sum_{j=0}^{q-2}\beta_{i_{j}}x^{j}$,
$\tilde{g}=\prod_{i=0}^{q-1}(g^{x^{i}})^{\alpha_{i}}=g^{f(x)}$,
$\hat{g}=\prod_{i=0}^{q-1}(g^{x^{i+1}})^{\alpha_{i}}=\tilde{g}^{x}$.
$\mathcal{B}$ selects $e,a,k\stackrel{R}{\leftarrow}\mathbb{Z}_{p}$ and computs
$h=((\hat{g}\tilde{g}^{e})^{k}\tilde{g}^{-1})^{\frac{1}{a}}= 
\tilde{g}^{\frac{(x+e)k-1}{a}}$.
 $\mathcal{B}$ selects  $x_{a},\gamma,\vartheta 
\stackrel{R}{\leftarrow}\mathbb{Z}_{p}$, and computes 
$Y_{A}=\mathfrak{g}^{x_{a}}$, $\xi=\tilde{g}^{\gamma}$ and 
$\tilde{h}=h^{\vartheta}$. $\mathcal{B}$ selects four three functions 
$H_{1}:\{0,1\}^{*}\rightarrow\mathbb{Z}_{p}$ and 
$H_{2}:\left\{0,1\right\}^{*}\rightarrow \mathbb{Z}_{p}$. $\mathcal{B}$ sends 
$(e,p,\mathbb{G}_{1},\mathbb{G}_{2},\mathbb{G}_{\tau},\tilde{g},h, 
\xi,\tilde{h},\mathfrak{g},Y_{A},H_{1},H_{2})$ to $\mathcal{A}$. \medskip

\noindent{\sf Registration Query.} $\mathcal{A}$ can  make the following 
queries. 

\begin{enumerate}%
\item{\sf Ticket Issuer Registration Query.} $\mathcal{B}$ sets
$\tilde{Y}_{I}=\mathfrak{g}^{x}$, and computes $Y_{I}=(\hat{g})^{\gamma}$.
$\mathcal{B}$ selects $e_{i},r_{i}\stackrel{R}{\leftarrow}\mathbb{Z}_{p}$ and
computes $\sigma_{I}=(\tilde{g}h^{r_{i}}Y_{I})^{\frac{1}{x_{a}+e_{i}}}$.
$\mathcal{B}$ sends $(Y_{I},\tilde{Y}_{I},\sigma_{I})$ to $\mathcal{A}$.
\medskip 

\item{\sf Ticket Verifier Registration Query.} $\mathcal{A}$ selects an identity
$ID_{V}$ and sends  $(ID_{V},Y_{V})$ to $\mathcal{B}$ where $Y_{V}$ is the
public key of $ID_{V}$. $\mathcal{B}$ selects
$x_{v},e_{v},r_{v}\stackrel{R}{\leftarrow}\mathbb{Z}_{p}$, and computes
$\sigma_{V}=(\tilde{g}h^{r_{v}}Y_{V})^{\frac{1}{x+e_{v}}}$. $\mathcal{B}$ sends
$( e_{v},r_{v},\sigma_{V})$ to $\mathcal{A}$. 
$\mathcal{A}$ can adaptively make this registration queries multiple times.
\medskip

\item{\sf User Registration Query.} $\mathcal{A}$ selects an identity $ID_{U}$
and sends $(ID_{U},Y_{U})$ to $\mathcal{B}$ where $Y_{U}$ is the public key of
$ID_{U}$.  $\mathcal{B}$ selects
$\lambda_{u},r_{u}\stackrel{R}{\leftarrow}\mathbb{Z}_{p}$, and computes
$\sigma_{U}=(\tilde{g}h^{r_{u}}Y_{U})^{\frac{1}{x+\lambda_{u}}}$. $\mathcal{B}$
sends $(\lambda_{u},r_{u},\sigma_{U})$ to $\mathcal{A}$. 
$\mathcal{A}$ can adaptively make this registration queries multiple times.
\medskip

\item{\sf Central Verifier Registration Query.} $\mathcal{A}$ selects an
identity $ID_{CV}$ and sends $(ID_{CV},Y_{CV})$ to $\mathcal{B}$ where $Y_{CV}$
is the public key of $ID_{CV}$. $\mathcal{B}$ selects
$\lambda_{cv},r_{cv}\stackrel{R}{\leftarrow}\mathbb{Z}_{p}$ and computes
$\sigma_{CV}=(\tilde{g}h^{r_{cv}}Y_{CV})^{\frac{1}{x_{a}+\lambda_{cv}}}$.
$\mathcal{B}$ sends $(\lambda_{cv},r_{cv},\sigma_{CV})$ to $\mathcal{A}$.
\end{enumerate}

\noindent{\sf Ticket Issuing Query.} $\mathcal{A}$ can adaptively submit a set
of {service} information  $J_{U}$, a set of pseudonyms
$Ps_{U}=\left\{(P_{V},Q_{V})_{ID_{V}\in J_{U}}\right\}$ and a proof
$\prod_{U}^{1}:~\mbox{PoK}\{(x_{u},r_{u},e_{u},\sigma_{U},v_{1},v_{2},v_{3},(z_{v})_{{V}\in
 J_{U}}): 
\frac{\tilde{\sigma}_{U}}{\bar{B}_{U}}=\bar{\sigma}_{U}^{-e_{u}}h^{v_{2}} 
\wedge g^{-1}=\bar{B}_{U}^{v_{3}}g^{x_{u}}h^{r_{u}-v_{2}v_{3}}\wedge(P_{V}= 
\xi^{x_{u}}{Y}_{P}^{z_{v}}\wedge Q_{V}=\xi^{z_{v}})_{{V}\in J_{U}}\}$. 
$\mathcal{B}$ verifies $\prod_{U}^{1}$ and 
$e(\bar{\sigma}_{U},Y)\stackrel{?}{=}e(\tilde{\sigma}_{U},\mathfrak{g})$. If 
the verification is unsuccessful, $\mathcal{B}$ aborts; otherwise, 
$\mathcal{B}$ works as follows. 
\medskip

\noindent For ${V}\in J_{U}$, let
$f_{v}(x)=\frac{f(x)}{x+e_{v}}=\sum_{k=1}^{q-2}\beta_{v_{k}}x^{k}$.
$\mathcal{B}$ selects $t_{u},d_{v},w_{v}\stackrel{R}{\leftarrow}\mathbb{Z}_{p}$ 
and   computes $C_{U}=g^{t_{u}}$,
$D_{V}=H_{2}(C_{U}||{V})$,~$E_{V}=\xi^{d_{v}}$, $F_{V}=Y_{V}^{d_{v}}$,
$K_{V}=Y_{V}{Y}_{P}^{d_{v}}$, $s_{v}=H_{1}(P_{V}||Q_{V}||E_{V}||F_{V}||K_{V}||$
$Text)$ and
$$Z_{V}=\prod_{k=0}^{q-2}(g^{x^{k}})^{\beta_{v_{k}}(1+\frac{(ek-1)s_{v}}{a})} 
\prod_{k=0}^{q-2}(g^{x^{k+1}})^{\frac{\beta_{v_{k}}ks_{v}}{a}}.$$
We claim that $(w_{v},e_{v},Z_{V})$ is a valid signature on $s_{v}$. We have
\begin{equation}%
\begin{split}%
Z_{V}
&=\prod_{k=0}^{q-2}(g^{x^{k}})^{\beta_{v_{k}}(1+\frac{(ek-1)(w_{v}+\vartheta
s_{v})}{a})}\prod_{k=0}^{q-2}(g^{x^{k+1}})^{\frac{\beta_{v_{k}}k(w_{v}+\vartheta
 s_{v})}{a}}\\
&=\prod_{k=0}^{q-2}(g^{\beta_{v_{k}}x^{k}})^{(1+\frac{(ek-1)(w_{v}+\vartheta
s_{v})}{a})}\prod_{k=0}^{q-2}(g^{\beta_{v_{k}}x^{k}})^{\frac{xk(w_{v}+\vartheta
 s_{v})}{a}}\\
&=(g^{\sum_{k=0}^{q-2}\beta_{v_{k}}x^{k}})^{(1+\frac{(ek-1)(w_{v}+\vartheta 
s_{v}}{a})}(g^{x\sum_{k=0}^{q-2}\beta_{v_{k}}x^{k}})^{\frac{k(w_{v}+\vartheta
 s_{v})}{a}}\\
& =(g^{f_{v}(x)})^{(1+\frac{(ek-1)(w_{v}+\vartheta 
s_{v})}{a})}(g^{xf_{v}(x)})^{\frac{k(w_{v}+\vartheta s_{v})}{a}}\\
&=(g^{f(x)})^{({1+\frac{(ek-1)(w_{v}+\vartheta 
s_{v})}{a})\frac{1}{x+e_{v}}}}(g^{xf(x)})^{\frac{k(w_{v}+\vartheta 
s_{v})}{a(x+e_{v})}}\\
&=\tilde{g}^{({1+\frac{(ek-1)(w_{v}+\vartheta 
s_{v})}{a})\frac{1}{x+e_{v}}}}\tilde{g}^{\frac{xk(w_{v}+\vartheta 
s_{v})}{a(x+e_{v})}}\\
&=(\tilde{g}^{({1+\frac{(ek-1)(w_{v}+\vartheta 
s_{v})}{a})}}\tilde{g}^{\frac{xk(w_{v}+\vartheta 
s_{v})}{a}})^{\frac{1}{x+e_{v}}}\\
&=(\tilde{g}\tilde{g}^{\frac{ekw_{v}}{a}}\tilde{g}^{\frac{-w_{v}}{a}}
\tilde{g}^{\frac{ek\vartheta s_{v}}{a}}\tilde{g}^{\frac{-\vartheta 
s_{v}}{a}}
\tilde{g}^{\frac{xkw_{v}}{a}}\tilde{g}^{\frac{xk\vartheta 
s_{v}}{a}})^{\frac{1}{x+e_{v}}}\\
&=\left(\tilde{g}\tilde{g}^{(\frac{(k(e+x)-1)w_{v}}{a}} 
\tilde{g}^{(\frac{(k(e+x)-1)\vartheta
 s_{v}}{a}}\right)^{\frac{1}{x+e_{v}}}\\
&=\left((\tilde{g}(\tilde{g}^{\frac{k(e+x)-1}{a}})^{w_{v}} 
((\tilde{g}^{\frac{(k(e+x)-1}{a}})^{\vartheta})^{s_{v}}\right)^{\frac{1}{x+e_{v}}}\\
&=(\tilde{g}h^{w_{v}}\tilde{h}^{s_{v}})^{\frac{1}{x+e_{v}}}
\end{split}%
\label{eq.1}
\end{equation}

Let $f_{cv}(x)=\frac{f(x)}{x+e_{cv}}=\sum_{k=0}^{q-2}\beta_{c_{k}}x^{k}$, where
$e_{cv}\in\{e_{1},e_{2},\cdots,e_{q-1}\}$.  $\mathcal{B}$ selects
$w_{cv}\stackrel{R}{\leftarrow}\mathbb{Z}_{p}$ and  computes
$s_{cv}=H_{1}(s_{1}||s_{2}||\cdots||s_{|J_{U}|})$ and
$Z_{CV}=\prod_{k=0}^{q-2}(g^{x^{k}})^{\beta_{c_{k}}(1+ 
\frac{(ek-1)(w_{cv}+\vartheta
s_{cv}}{a})}\prod_{k=0}^{q-2}(g^{x^{k+1}})^{\frac{\beta_{c_{k}}k(w_{cv}+\vartheta
 s_{cv})}{a}}.$

According to Equation (\ref{eq.1}), $(w_{cv},e_{cv},Z_{CV})$ is a BBS+ signature
on $s_{cv}$.
	
If the $q$-th signature is required, $\mathcal{B}$ computes 
$w_{cv}=a-\vartheta s_{u}$ and $Z_{CV}=\tilde{g}^{k}$.   We claim that 
$(w_{cv},e,Z_{CV})$ is valid signature on $s_{cv}$. Because, we have
\begin{equation*}%
\begin{split}
Z_{CV}&=\tilde{g}^{k}= 
(\tilde{g}\tilde{g}^{\frac{a(k(x+e)-1)}{a}})^{\frac{1}{x+e}}= 
(\tilde{g}\tilde{g}^{\frac{(w_{cv}+\vartheta
s_{cv})(k(x+e)-1)}{a}})^{\frac{1}{x+e}}= 
(\tilde{g}\tilde{g}^{\frac{w_{cv}(k(x+e)-1)}{a}}\tilde{g}^{\frac{\vartheta
s_{p}(k(x+e)-1)}{a}})^{\frac{1}{x+e}}\\
&=\left(\tilde{g}(\tilde{g}^{\frac{k(x+e)-1}{a}})^{w_{cv}} 
((\tilde{g}^{\frac{k(x+e)-1}{a}})^{\vartheta})^{s_{cv}}\right)^{\frac{1}{x+e}}= 
(\tilde{g}h^{w_{cv}}\tilde{h}^{s_{cv}})^{\frac{1}{x+e}}.
\end{split}
\end{equation*}

The ticket is $T_{U}=\big\{(D_{V},Tag_{V})|{V}\in J_{U}\big\} \cup$
\big\{$s_{cv},w_{cv},e_{cv},Z_{CV}\big\}$. $\mathcal{B}$ sends $(C_{U},$
$T_{U},Text)$ to $\mathcal{A}$. Let $QT$ be the set consisting of the tickets
queried by $\mathcal{A}$ and is initially empty.  $\mathcal{B}$ adds  
$(T_{U},C_{U})$ into $QT$. $\mathcal{A}$ can make this query adaptively at most
$\varrho\leq q$ times. \medskip
	
\noindent{\sf Output:} $\mathcal{A}$ outputs a ticket
$T_{U^{*}}=\big\{(D_{V^{*}},P_{V^{*}},Q_{V^{*}},E_{V^{*}},F_{V^{*}}, K_{V^{*}},
s_{v^{*}},w_{v^{*}},e_{v^{*}},Z_{V^{*}},Text)|{V^{*}}\in J_{U^{*}}\big\} \cup~
\big\{(s_{cv},w_{cv},e_{cv},Z_{CV})\big\}$. Let $(s^{*},w^{*},e^{*},Z^{*})\in
\big\{(s_{v^{*}},w_{v^{*}},e_{v^{*}},Z_{V^{*}}|{V^{*}}\in J_{U^{*}}\big\}\cup
\big\{(s_{cv},w_{cv},e_{cv},$\\$Z_{CV})\big\}$ be a forged authentication tag.
\medskip
	
We consider the following three cases.

\begin{itemize}%
\item{\sf Case-I.} $e^{*}\notin\{e_{1},e_{2},\cdots,e_{q-1},e\}$.  Let
$f_{1}^{*}(x)=\frac{f(x)}{x+e^{*}}=\sum_{i=0}^{q-2}c_{i}x^{i}$,
$f_{2}^{*}(x)=\frac{f(x)(e+x)}{x+e^{*}}=\sum_{i=0}^{q-1}\tilde{c}_{i}x^{i}$ and
$f(x)=(x+e^{*})c(x)+\theta_{0}$ where $c(x)=\sum_{i=0}^{q-2}c_{i}x^{i}$.
Therefore,
\begin{equation*}%
\sigma^{*}=(\tilde{g}h^{w^{*}}\tilde{h}^{s^{*}})^{\frac{1}{x+e^{*}}}= 
\tilde{g}^{\frac{1}{x+{e^{*}}}}(h^{w^{*}}\tilde{h}^{s^{*}})^{\frac{1}{x+e^{*}}}.
\end{equation*} 

We have
\begin{equation*}%
\begin{split}%
\tilde{g}^{\frac{1}{x+e^{*}}}&=\sigma^{*}\cdot
(h^{w^{*}}\tilde{h}^{s^{*}})^{\frac{-1}{x+e^{*}}}\\ &=\sigma^{*}\cdot
(\tilde{g}^{\frac{w^{*}((e+x)-1)}{a}}\tilde{g}^{\frac{\vartheta
s^{*}((e+x)-1)}{a}})^{\frac{-1}{x+e^{*}}}\\ 
&=\sigma^{*}\cdot \tilde{g}^{\frac{-(w^{*}+ \vartheta
s^{*})(x+e)}{a(x+e^{*})}}\cdot \tilde{g}^{\frac{w^{*}+ \vartheta
s^{*}}{a(x+e^{*})}}\\ & =\sigma^{*}\cdot 
{g}^{\frac{-f(x)(w^{*}+\vartheta
s^{*})(x+e)}{a(x+e^{*})}}\cdot {g}^{\frac{f(x)(w^{*}+\vartheta
s^{*})}{a(x+e^{*})}}\\ &= \sigma^{*}\cdot {g}^{\frac{-(w^{*}+\vartheta
s^{*})f_{2}^{*}(x)}{a}}\cdot {g}^{\frac{(w^{*}+\vartheta
s^{*})f_{1}^{*}(x)}{a}}\\ & =\sigma^{*}\cdot
\prod_{k=0}^{q-1}(g^{x^{k}})^{\frac{-\tilde{c}_{k}(w^{*}+\vartheta
s^{*})}{a}}\cdot 
\prod_{k=0}^{q-2}(g^{x^{k}})^{\frac{c_{i}(w^{*}+\vartheta
s^{*})}{a}}.
\end{split}
\end{equation*}
		
Let $\Gamma=\sigma^{*}\cdot 
\prod_{k=0}^{q-1}(g^{x^{k}})^{\frac{-\tilde{c}_{k}(w^{*}+\vartheta 
s^{*})}{a}}\cdot  (g^{x^{i}})^{\frac{c_{k}(w^{*}+\vartheta 
s^{*})}{a}}$.  We have 

\begin{equation*}%
\Gamma=\tilde{g}^{\frac{1}{x+e^{*}}}=g^{\frac{f(x)}{x+e^{*}}}= 
g^{\frac{c(x)(x+e^{*})+\theta}{x+e^{*}}}=g^{c(x)}g^{\frac{\theta_{0}}{x+e^{*}}}. 
\end{equation*}%
Hence, 
\begin{equation*}%
 g^{\frac{1}{x+e^{*}}}=(\Gamma\cdot 
g^{-c(x)})^{\frac{1}{\theta}}=\left(\sigma^{*}\cdot 
\prod_{k=0}^{q-1}(g^{x^{i}})^{\frac{-\tilde{c}_{k}(w^{*}+\vartheta 
s^{*})}{a}}\cdot  
\prod_{k=0}^{q-2}(g^{x^{i}})^{\frac{c_{k}(w^{*}+\vartheta 
s^{*})}{a}}\cdot 
\prod_{k=0}^{q-2}(g^{x^{k}})^{-c_{k}}\right)^{\frac{1}{\theta}}.
\end{equation*}
\medskip

\item{\sf Case-II.} $e^{*}\in\{e_{1},e_{2},\cdots,e_{q-1},e\}$. We have 
$e^{*}=e$ with the probability $\frac{1}{q}$. Since 
$e\notin\{e_{1},e_{2},\cdots,$ $e_{q-1}\}$, $\mathcal{B}$ can output 
$g^{\frac{1}{x+e}}$ using the same technique above. 
\medskip

\item{\sf Case-III.} $e^{*}=e_{v}$, $\sigma^{*}=\sigma_{V}$, but 
$s^{*}\neq s_{v}$. Since 
$\sigma^{*}=(\tilde{g}h^{w^{*}}\tilde{h}^{s^{*}})^{\frac{1}{x+e^{*}}}$ 
and 
$\sigma_{v}=(\tilde{g}h^{w_{V}}\tilde{h}^{s_{v}})^{\frac{1}{x+e_{v}}}$. 
We have $h^{w^{*}}\tilde{h}^{s^{*}}=h^{w_{v}}\tilde{h}^{s_{v}}$, 
$\tilde{h}=h^{\frac{w^{*}-w_{v}}{s_{v}-s^{*}}}$ and 
$log_{h}\tilde{h}={\frac{w^{*}-w_{v}}{s_{v}-s^{*}}}$. $\mathcal{B}$ can 
use $\mathcal{A}$ to break the discrete logarithm assumption. Therefore 
$\mathcal{B}$ can use $\mathcal{A}$ to break the JOC-$q$-SDH assumption 
since JOC-$q$-SDH
assumption is included in discrete logarithm assumption.
\end{itemize}
	
Therefore, the advantage with which $\mathcal{B}$ can break the $q$-SDH 
assumption is 
\begin{equation*}%
\begin{split}%
Adv_{\mathcal{B}}^{q-SDH}=&\Pr[{\sf Case-I}]+\Pr[{\sf 
Case-II}]+\Pr[{\sf Case-III}]\\
& \geq\frac{p-q}{p}\epsilon'(\ell)+\frac{q}{p}\times\frac{1}{q} 
\epsilon'(\ell)+\frac{1}{p}\times\frac{1}{p}\times
 \frac{p-1}{p}\epsilon'(\ell)\\
& =(\frac{p-q}{p}+\frac{1}{p}+\frac{p-1}{p^{3}})\epsilon'(\ell).
\end{split}
\end{equation*}

\end{proof}

\section{Proof of Theorem \ref{theorem:trace}}%
\label{app:trace}
\begin{proof}
If there exists an adversary $\mathcal{A}$ which can break the traceability of our  scheme with the advantage $\epsilon'(\ell)$, we can constructs an algorithm $\mathcal{B}$ which can use $\mathcal{A}$ as a subroutine to break the JOC-$q$-SDH assumption as follows. Given a $(q+3)$-tuple $(g,g^{x},\cdots,g^{x^{q}},\mathfrak{g},\mathfrak{g}^{x})$, $\mathcal{B}$ will output $(c,g^{\frac{1}{x+c}})$ where $c\in\mathbb{Z}_{p}-\{-x\}$.
\medskip\\
\noindent{\sf Setup.}  $\mathcal{B}$ selects $e_{1},e_{2},\cdots,e_{q-1}\stackrel{R}{\leftarrow}\mathbb{Z}_{p}$, and sets $f(x)=\prod_{i=1}^{q-1}(x+e_{i})=\sum_{i=0}^{q-1}\alpha_{i}x^{i}$, $f_{i}(x)=\frac{f(x)}{x+e_{i}}=\sum_{j=0}^{q-2}\beta_{i_{j}}x^{j}$, $\tilde{g}=\prod_{i=0}^{q-1}(g^{x^{i}})^{\alpha_{i}}=g^{f(x)}$, $\hat{g}=\prod_{i=0}^{q-1}(g^{x^{i+1}})^{\alpha_{i}}=\tilde{g}^{x}$.  $\mathcal{B}$ selects $e,a,k\stackrel{R}{\leftarrow}\mathbb{Z}_{p}$ and computs $h=((\hat{g}\tilde{g}^{e})^{k}\tilde{g}^{-1})^{\frac{1}{a}}=\tilde{g}^{\frac{(x+e)k-1}{a}}$. $\mathcal{B}$ selects  $x_{a},\gamma,\vartheta \stackrel{R}{\leftarrow}\mathbb{Z}_{p}$, and computes $Y_{A}=\mathfrak{g}^{x_{a}}$, $\xi=\tilde{g}^{\gamma}$ and $\tilde{h}=h^{\vartheta}$. $\mathcal{B}$ selects four three functions $H_{1}:\{0,1\}^{*}\rightarrow\mathbb{Z}_{p}$ and  $H_{2}:\left\{0,1\right\}^{*}\rightarrow \mathbb{Z}_{p}$.
$\mathcal{B}$ sends 
$(e,p,\mathbb{G}_{1},\mathbb{G}_{2},\mathbb{G}_{\tau},\tilde{g},h,\xi, 
\tilde{h},\mathfrak{g},Y_{A},H_{1},H_{2})$ to $\mathcal{A}$.
\medskip\\
\noindent{\sf Registration Query.} $\mathcal{A}$ can  make the following queries. 
\begin{enumerate}
\item{\sf Ticket Issuer Registration Query.} $\mathcal{B}$ sets $\tilde{Y}_{I}=\mathfrak{g}^{x}$, and computes $Y_{I}=(\hat{g})^{\gamma}$. $\mathcal{B}$ selects $e_{i},r_{i}\stackrel{R}{\leftarrow}\mathbb{Z}_{p}$ and computes $\sigma_{I}=(\tilde{g}h^{r_{i}}Y_{I})^{\frac{1}{x_{a}+e_{i}}}$. $\mathcal{B}$ sends $(Y_{I},\tilde{Y}_{I},\sigma_{I})$ to $\mathcal{A}$.
\medskip
\item{\sf Ticket Verifier Registration Query.} $\mathcal{A}$ selects an identity $ID_{V}$ and sends  $(ID_{V},Y_{V})$ to $\mathcal{B}$ where $Y_{V}$ is the public key of $ID_{V}$. $\mathcal{B}$ selects $x_{v},e_{v},r_{v}\stackrel{R}{\leftarrow}\mathbb{Z}_{p}$, and computes $\sigma_{V}=(\tilde{g}h^{r_{v}}Y_{V})^{\frac{1}{x+e_{v}}}$.  $\mathcal{B}$ sends $( e_{v},r_{v},\sigma_{V})$ to $\mathcal{A}$.  
$\mathcal{A}$ can adaptively make this registration queries multiple times.
\medskip
\item{\sf User Registration Query.} $\mathcal{A}$ selects an identity $ID_{U}$ and sends $(ID_{U},Y_{U})$ to $\mathcal{B}$ where $Y_{U}$ is the public key of $ID_{U}$.  $\mathcal{B}$ selects $\lambda_{u},r_{u}\stackrel{R}{\leftarrow}\mathbb{Z}_{p}$, and computes $\sigma_{U}=(\tilde{g}h^{r_{u}}Y_{U})^{\frac{1}{x+\lambda_{u}}}$. $\mathcal{B}$ sends $(\lambda_{u},r_{u},\sigma_{U})$ to $\mathcal{A}$.   
$\mathcal{A}$ can adaptively make this registration queries multiple times.
\medskip
\item{\sf Central Verifier Registration Query.} $\mathcal{A}$ selects an identity $ID_{CV}$ and sends $(ID_{CV},Y_{CV})$ to $\mathcal{B}$ where $Y_{CV}$ is the public key of $ID_{CV}$. $\mathcal{B}$ selects $\lambda_{cv},r_{cv}\stackrel{R}{\leftarrow}\mathbb{Z}_{p}$ and computes $\sigma_{CV}=(\tilde{g}h^{r_{cv}}Y_{CV})^{\frac{1}{x_{a}+\lambda_{cv}}}$. $\mathcal{B}$ sends $(\lambda_{cv},r_{cv},\sigma_{CV})$ to $\mathcal{A}$. 
\end{enumerate}
\medskip
\noindent{\sf Ticket Issuing Query.} $\mathcal{A}$ can adaptively submit a set 
of {service} information  $J_{U}$, a set of pseudonyms 
$Ps_{U}=\left\{(P_{V},Q_{V})_{ID_{V}\in J_{U}}\right\}$ and a proof 
$\prod_{U}^{1}:~\mbox{PoK}\{(x_{u},r_{u},e_{u},\sigma_{U},v_{1},v_{2},v_{3},(z_{v})_{{V}\in
 J_{U}}): 
\frac{\tilde{\sigma}_{U}}{\bar{B}_{U}}=\bar{\sigma}_{U}^{-e_{u}}h^{v_{2}} 
\wedge g^{-1}=\bar{B}_{U}^{v_{3}}g^{x_{u}}h^{r_{u}-v_{2}v_{3}}\wedge(P_{V}= 
\xi^{x_{u}}{Y}_{P}^{z_{v}}\wedge Q_{V}=\xi^{z_{v}})_{{V}\in J_{U}}\}$. 
$\mathcal{B}$ verifies $\prod_{U}^{1}$ and 
$e(\bar{\sigma}_{U},Y)\stackrel{?}{=}e(\tilde{\sigma}_{U},\mathfrak{g})$. If 
the verification is unsuccessful, $\mathcal{B}$ aborts; otherwise, 
$\mathcal{B}$ generates a ticket $T_{U}=\big\{(D_{V},$ $Tag_{V})|{V}\in 
J_{U}\big\} \cup$ \big\{$s_{cv},w_{cv},e_{cv},Z_{CV}\big\}$ by using the 
technique gaven in the proof of Theorem 1, and sends $T_{U}$ to $\mathcal{A}$.  
Let $QT$ be the set consisting of the tickets queried by $\mathcal{A}$ and is 
initially empty.  $\mathcal{B}$ adds   $(T_{U},C_{U})$ into $QT$. $\mathcal{A}$ 
can make this query adaptively at most $\varrho\leq q$ times.
\medskip\\

\noindent{\sf Output:} $\mathcal{A}$ outputs a ticket 
$T_{U^{*}}=\big\{(D_{V^{*}},P_{V^{*}},Q_{V^{*}},E_{V^{*}},F_{V^{*}}, K_{V^{*}}, 
s_{v^{*}},w_{v^{*}},e_{v^{*}},Z_{V^{*}},Text)|{V^{*}}\in J_{U^{*}}\big\} \cup~ 
\big\{(s_{cv},w_{cv},e_{cv},Z_{CV})\big\}$.  If more than one users' pubic keys 
are included in the ticket $T_{U^{*}}$, the ticket is not generated correctly 
and $\mathcal{B}$ aborts. If $\mathcal{B}$ does not abort,  the following two 
types of forgers are considered. Type-I forgers outputs a ticket $T_{U}^{*}$  
which includes at least a new pseudonym $ (P'_{V},Q'_{V})$  which is not 
included in any ticket queried by $\mathcal{A}$. Type-II forger outputs a 
ticket $T_{U}^{*}$ which includes the same pseudonyms  included in a ticket 
$T_{U}\in QT$ queried by $\mathcal{A}$, but can be trace to a user 
$\mathcal{U}'$ whose secrete key $x'$ is not known by $\mathcal{A}$. Let 
$(x',Y')$ be the secret-public key pair of $\mathcal{U}'$.

\begin{itemize}

\item{\em Type-I:} If there is a pseudonym $(P'_{V},Q'_{V})\subset Tag'_{V}\in 
T_{U^{*}}$ and $(P'_{V},Q'_{V})\notin QT$.  Let 
$Tag'_{V}=(P'_{V},Q'_{V},E'_{V},$ 
$F'_{V},K'_{V},Z'_{V},s'_{v},w'_{v},e'_{v},Text)$. $\mathcal{A}$  forged a 
signature $(w'_{v},e'_{v},Z'_{V})$ on $s'_{v}$ where 
$s'_{v}=H_{1}(P'_{V}||Q'_{V}||E'_{V}||F'_{V}||K'_{V}||Text)$. Hence, 
$\mathcal{B}$ can use $\mathcal{A}$ to break the JOC-version-q-SDH assumption 
by using the technique in the proof of Theorem 1.

\item{\em Type-II.} If all pseudonyms $(P_{V},Q_{V})\subset Tag_{V}\in 
T_{U^{*}}$ and $(P_{V},Q_{V})\in QT$. In $T_{U}^{*}$, there is a pseudonym 
$(P_{CV},Q_{CV})$ generated for the central verifier. If $\mathcal{A}$ can 
generate a proof  $\mbox{PoK}\{(x',z_{cv}): 
P_{CV}=\xi^{x'}{Y}_{CV}^{z_{cv}}~\wedge~Q_{V}=\xi^{z_{cv}}\}$, $\mathcal{B}$ 
can use the rewinding technique to extract the knowledge of $(x',z_{cv})$ from 
$\mathcal{A}$, namely given $(\xi,Y')$, $\mathcal{B}$ can output a $x'$ such 
that $Y'=\xi^{x'}$. Hence, $\mathcal{B}$ can use $\mathcal{A}$ to break the 
discrete logarithm assumption.

\end{itemize}

Let $\Pr[\mbox{Type-I}]$ and $\Pr[\mbox{Type-II}]$ denote the probabilities 
with which $\mathcal{A}$ can success, respectively.
By Theorem \ref{theorem:unforge}, we have $\Pr[\mbox{Type-I}]=\frac{1}{2}\times 
\epsilon_{1}(\ell)(\frac{p-q}{p}+\frac{1}{p}+\frac{p-1}{p^{3}})$.  Hence, 
$\mathcal{B}$ can break the $q$-SDH assumption with the advantage 
$\frac{\epsilon_{1}(\ell)}{2}(\frac{p-q}{p}+\frac{1}{p}+\frac{p-1}{p^{3}})$ or 
break the DL assumption with the advantage $\frac{\epsilon_{2}(\ell)}{2}$. 
Therefore, 
$\epsilon(\ell)=max\left\{\frac{\epsilon_{1}(\ell)}{2}(\frac{p-q}{p}+ 
\frac{1}{p}+\frac{p-1}{p^{3}}),\frac{\epsilon_{2}(\ell)}{2}\right\}$.
\end{proof}

\end{document}